\documentclass[11pt,twoside,fleqn]{article}
\usepackage{color,graphicx,lscape,here,geometry,setspace,multirow,enumitem}
\usepackage[dvipsnames]{xcolor}
\usepackage{graphicx}
\usepackage{verbatim}

\usepackage{natbib}
\usepackage{bibunits}
\defaultbibliography{references}
\defaultbibliographystyle{custom}

\usepackage{amsfonts}
\usepackage{authblk}
\usepackage{silence}

\usepackage[misc,geometry]{ifsym}

\usepackage{placeins}
\usepackage{booktabs}
\usepackage{color,colortbl}
\usepackage{setspace}
\usepackage{enumitem}
\usepackage{amsfonts,amsmath,amssymb}

\newlist{steps}{enumerate}{1}
\setlist[steps,1]{wide=0pt, leftmargin=\parindent, label=Step \arabic*:, font=\scshape}

\definecolor{nblue}{HTML}{000660}
\usepackage[colorlinks=true,urlcolor=nblue,linkcolor=nblue,citecolor=nblue]{hyperref}

\newcommand{\diag}{\text{diag}}

\usepackage{tabularx}
\usepackage{threeparttable}
\usepackage{dcolumn}
\newcolumntype{d}[1]{D{.}{.}{#1}}

\usepackage{array}
\newcolumntype{C}[1]{>{\centering\arraybackslash}p{#1}}

\usepackage[hang,flushmargin]{footmisc} 
\usepackage[hang]{footmisc}
\usepackage{float}
\restylefloat{table}

\usepackage[title,titletoc]{appendix}
\makeatletter

\renewenvironment{appendices}{%
    \begin{oldappendices}%
    \renewcommand{\thefigure}{\ifnum \c@section>\z@ \thesection.\fi\@arabic\c@figure}%
    \@addtoreset{figure}{section}%
    \renewcommand{\thetable}{\ifnum \c@section>\z@ \thesection.\fi\@arabic\c@table}%
    \@addtoreset{table}{section}}{%
    \end{oldappendices}%
}\makeatother

\let\emptyset\varnothing

\usepackage{titlesec} 
\titleformat{\section}[block]{\bfseries\sffamily\Large}{\thesection. }{0em}{} 
\titleformat{\subsection}[block]{\bfseries\sffamily\large}{\thesubsection. }{0em}{} 
\titleformat{\subsubsection}[block]{\large}{}{0em}{\itshape} 

\newcommand{\bi}{\begin{itemize}}
\newcommand{\ei}{\end{itemize}}
\newcommand{\be}{\begin{equation}}
\newcommand{\ee}{\end{equation}}
\defcitealias{ieo14}{IEO, 2014}

\long\def\symbolfootnote[#1]#2{\begingroup%
\def\thefootnote{\fnsymbol{footnote}}\footnote[#1]{#2}\endgroup}

\makeatletter
\def\ubar#1{\underline{\sbox\tw@{$#1$}\dp\tw@\z@\box\tw@}}
\def\obar#1{\overline{\sbox\tw@{$#1$}\dp\tw@\z@\box\tw@}}
\makeatother

\usepackage{bm}
\usepackage{caption}
\usepackage{subcaption}

\makeatletter\let\p@subfigure\thefigure\makeatother

\floatstyle{plaintop}
\restylefloat{table}

\usepackage{fancyhdr} 
\usepackage{cleveref}
\crefname{chapter}{Chapter}{Chapters}
\crefname{section}{Section}{Sections}
\crefname{subsection}{Section}{Sections}
\crefname{subsubsection}{Section}{Sections}
\crefname{figure}{Figure}{Figures}
\crefname{table}{Table}{Tables}
\crefname{equation}{Equation}{Equations}
\crefname{appendix}{Appendix}{Appendices}
\crefname{appendices}{Appendix}{Appendices}

\crefname{appsec}{Appendix}{Appendices}

\def\titletext{General Seemingly Unrelated Local Projections}

\title{\sffamily\huge{\textbf{\titletext}}}
\author{}
\date{}

\usepackage[utf8, latin1]{inputenc}                 

\begin{document}

\maketitle
\vspace*{-6em} 
\begin{center}

\end{center}
\begin{center}
\begin{minipage}{.32\textwidth}
  \centering\normalsize Florian \MakeUppercase{Huber}\\[0.25em]
  \small \textit{University of Salzburg}
\end{minipage}
\begin{minipage}{.32\textwidth}
  \centering\normalsize Christian \MakeUppercase{Matthes}\\[0.25em]
  \small \textit{University of Notre Dame}
\end{minipage}
\begin{minipage}{.32\textwidth}
  \centering\normalsize Michael \MakeUppercase{Pfarrhofer}\\[0.25em]
  \small \textit{WU Vienna}
\end{minipage}
\end{center}

\vspace*{1em}
\doublespacing
\begin{center}
\begin{minipage}{0.85\textwidth}
\noindent\small We develop a flexible framework for  Bayesian estimation of impulse responses using Local Projections (LPs) with instrumental variables. It accommodates multiple shocks and instruments, accounts for autocorrelation in multi-step forecasts by jointly modeling all LPs as a seemingly unrelated system of equations, defines a flexible yet parsimonious joint prior for impulse responses based on a Gaussian Process, and allows for joint inference about the entire vector of impulse responses. We show via Monte Carlo simulations that our approach delivers more accurate point and uncertainty estimates than standard methods. To address misspecification, we propose an optional robustification step based on power posteriors.\\
\textbf{\sffamily JEL}: C11, C22, C26, E00 \\
\textbf{\sffamily KEYWORDS}: Local Projections, Impulse Responses, Instruments, Bayesian Methods, Gaussian Process
\end{minipage}
\end{center}

\singlespacing\vfill\noindent{\footnotesize
We would like to thank Gary Koop and seminar participants at the Bank of Canada, the University of Strathclyde and the 15th RCEA Bayesian Econometrics Workshop for useful comments.}

\thispagestyle{empty}\renewcommand{\footnotelayout}{\setstretch{1}}
\doublespacing\normalsize\renewcommand{\thepage}{\arabic{page}}
\newpage\begin{bibunit}

\section{Introduction}\label{sec:introduction}
Impulse response functions (IRFs) represent the difference between forecasts conditioned on different values of specific shocks. In analogy to the forecasting literature, in a seminal contribution, \cite{jorda2005estimation} proposed Local Projections (LPs) as impulse response estimators akin to direct forecasts, in contrast to Vector Autoregressions \citep[VARs,][]{Sims} that imply estimates of impulse responses based on iterated forecasts. The simplicity of LPs --- they can be estimated via Ordinary Least Squares (OLS) and adjusted standard errors --- combined with extensions such as the use of instruments \citep{SW_EJ}, has made them arguably the go-to method for estimating impulse responses in macroeconomics and related fields.\footnote{LPs have also begun to make inroads in applied micro --- see \cite{Dube}.}

In finite samples, there is a (bias--variance) trade-off between the simplicity of LPs and the structure that VARs and related models impose \citep{li2024local}.\footnote{Asymptotically, both LPs and VARs estimate the same IRFs \citep{plagborg2021local}. Additional discussions on the relationship between VAR-based and LP-based inference about IRFs are provided in \citet{ludwig2024local} and \citet{baumeister2025NBERMA}.} While more robust against misspecification, LP estimators are typically less efficient, and the estimated impulse responses can behave erratically since horizon-specific LPs are treated independently. In such cases, some form of regularization is generally beneficial and can be used to introduce prior information about the shape of the IRFs. Furthermore, heteroskedasticity and autocorrelation consistent (HAC) standard errors do not exploit the available information about the correlation structure of the forecast errors across horizons. 

In this paper, we present a general framework to tackle these issues jointly while also confronting issues that are often set aside in LPs, such as the joint estimation of impulse responses to various shocks, in which case we need to tackle the issue that instruments for different shocks may be correlated in practice, as we show in one application. In that case, researchers need to take a stand on whether the correlation stems from noise (and needs to be filtered out) or if there is a common component/shock that itself should be studied. We investigate both alternatives. In addition, we explore how estimating LPs as a system of seemingly unrelated regressions takes the correlation of forecast errors into account.

This joint modeling enables us to design Bayesian machine learning priors on the joint distribution of the impulse responses. These are based on Gaussian processes (GPs), a nonparametric method popular in statistics and computer science,  that can be used to introduce information on the shape and magnitudes of the dynamic responses.  To decide on the importance of this information, we use Bayesian shrinkage and regularization techniques that researchers in macroeconomics are familiar with.

Our approach constructs a pseudo-/quasi-likelihood: a likelihood function that need not correspond to the process that generated the data, but can still provide useful guidance for estimation \citep{fiorentiniSentana2023}.\footnote{We explain the use of a \emph{pseudo}-likelihood in Section~\ref{sec:lp-ar}.} The frequentist and Bayesian misspecification results of \citet{white1982maximum} and \citet{fernandezVillaverdeRubioRamirez2004} show that likelihood-based estimators target pseudo-true values asymptotically: the values within the maintained misspecified model that best approximate the data-generating process under the relevant Kullback--Leibler criterion. Our setup makes it straightforward to construct error bands and related objects using the output of our posterior sampler---an area where other regularization methods often face complications. Because this pseudo-likelihood is misspecified, we address its use directly in two ways. First, we estimate the local projections jointly as a system and impose a flexible nonparametric prior. As we show in Section~\ref{sec: monte_carlo}, this leads to improved coverage probabilities relative to standard local projection specifications. Second, to further guard against the misspecification of the pseudo-likelihood, we introduce a novel, optional post-processing step that aligns the posterior with a power posterior \citep[see, e.g.,][]{holmes2017assigning,grunwald2017inconsistency}. Fractional and power posterior results provide formal support for likelihood downweighting under misspecification \citep{bhattacharya2019bayesian}.

Bayesian inference in high-dimensional parameter spaces can seem daunting to researchers used to the ease of OLS-based estimation of LPs. To help in that regard, we use a hierarchical model so that only very few prior hyperparameters need to be set. And for those hyperparameters, we offer benchmark values. Our approach estimates LPs on standardized data (before translating coefficients back to the original scale if desired), so the prior can be used across applications easily.

We are not the first to tackle some of these issues --- other papers have tackled subsets of the issues we have discussed so far. The original \cite{jorda2005estimation} paper discusses system estimation (in a Generalized Method of Moments context). \cite{lusompa2023local} derives the moving average (MA) structure of the forecast error in LPs and describes frequentist generalized least squares estimators to take this structure into account.\footnote{A similar approach is used in a nonlinear setting by \cite{Mumtaz}.} 

There are several important contributions dealing with regularization in LPs. In a frequentist context, \cite{barnichon2019impulse} use splines to achieve regularization. Closely related in the forecasting literature, \cite{baek2019forecasting} studies smoothed direct forecasts at long horizons. \cite{barnichon2019impulse} note the aforementioned conceptual difficulty in constructing error bands in this context. \cite{ferreira2023bayesian} instead use a Bayesian approach for LPs, but estimate the model one horizon at a time without directly taking into account the correlation of the forecast error, requiring them to ex-post adjust standard errors using a frequentist, HAC-based, approach. \cite{tanaka2020bayesian} also relies on Bayesian inference, and, to define a (joint) roughness penalty prior on the impulse response vector, estimates a system of LPs as a whole. Neither of these latter papers, however, takes into account instruments, one component of the framework we propose. \cite{AruobaDrechselJME} also estimate a system of local projections, but stack equations \emph{across variables}, not across horizons.\footnote{They mention that stacking across horizons is also possible, but do not estimate a system of local projections across variables and horizons due to the implied computational burden.}

Since we jointly estimate all impulse responses across horizons (and in some cases also across various shocks), our approach automatically allows for joint inference on impulse responses, a topic that has recently received substantial attention both in the VAR \citep{inoue2022} and LP \citep{inoue2024inference} literature.\footnote{Some papers estimate impulse responses directly using moving average models where joint inference is also natural \citep{barnichon2018functional, plagborg2019bayesian}.} In the applications below we report pointwise summaries; posterior draws from the full system can also be used to summarize functionals of the impulse-response path. More generally, our approach offers an alternative to recent frequentist approaches to inference in LPs \citep{montiel2021local,Xu}.

Finally, our framework is set up to deal with missing values. LPs often require researchers to drop some data or use different sample sizes for different horizons because of their structure as estimated direct forecasts. We instead impute missing values within a unified framework, allowing researchers to use the full sample across all horizons.

The remainder of the paper is structured as follows. Section \ref{sec:econometrics} introduces our econometric framework, whereas Section \ref{sec: general} offers a discussion on model extensions to showcase the generality of our approach. We illustrate the model based on synthetic and real data in Sections \ref{sec: monte_carlo} and \ref{sec: applications} and offer a summary and conclusions in Section \ref{sec:conclusions}.

\section{Econometric Framework}\label{sec:econometrics}
\subsection{Local projections as seemingly unrelated regressions}\label{sec:lp-ar}
This section begins by discussing the relationship between dynamic models and LPs using a simple example. It motivates our general modeling approach, which we describe in more detail in the next section. Let $w_t$ denote our variable of interest, which follows an AR(1) process:
\begin{equation*}
    w_{t} = \rho w_{t-1} + \varepsilon_{t},
\end{equation*}
where $\rho \in [-1, 1]$ denotes the autoregressive coefficient and $\varepsilon_{t}$ denotes an independently and identically distributed (iid) shock with variance $\sigma_\varepsilon^2$. These assumptions serve two purposes: they are used to justify (i) the control variables (so that the forecast errors are stationary), and (ii) the iid assumption on the forecast error for horizon 0. However, these high-level assumptions (stationary forecast errors, iid forecast errors at horizon 0) could be justified by alternative low-level assumptions such as stationary data \citep{lusompa2023local} so that we can work with a Wold representation. We could also allow for a possibly non-stationary AR($p$) process if we control for $p$ lags of the dependent variable.

We can iterate the equation above forward to retrieve:
\begin{equation*}
    w_{t+h} = \rho^{h+1} w_{t-1} + \rho^{h} \varepsilon_{t} + \dots + \rho \varepsilon_{t+h-1} + \varepsilon_{t+h}.
\end{equation*}
For $h=0, \dots, \tilde{H},$ we can represent these projections as a system of $H = \tilde{H}+1$ equations:
\begin{align}
    w_{t} = b_0 w_{t-1} + u_{t}^{(0)} \quad \hdots \quad w_{t+\tilde{H}} = b_h w_{t-1} + u_{t+\tilde{H}}^{(\tilde{H})},\label{eq: system-of-lps}
\end{align}
where $b_h = \rho^{h+1}$ denotes the dynamic multiplier and the shocks to the $h$-specific equation can be linked to the AR shocks by noting that:
\begin{equation*}
    u_{t+h}^{(h)} = \rho^{h} \varepsilon_{t} + \rho^{h-1} \varepsilon_{t+1} + \hdots + \rho \varepsilon_{t+h-1} + \varepsilon_{t+h}.
\end{equation*}
Equation (\ref{eq: system-of-lps}) illustrates that one may regress $w_{t+h}$ on $w_{t-1}$ to obtain direct estimates of dynamic multipliers (since we want to illustrate the benefits of a system approach in this Section, we do not explicitly introduce instruments for structural shocks yet). 

These are LPs in the spirit of \cite{jorda2005estimation}. However, doing so by OLS neglects the autocorrelation in forecast errors, which follow a moving average process of order $h$. As mentioned above, \cite{lusompa2023local} shows that the MA($h$) structure of the forecast error can be obtained under mild regularity conditions and does \emph{not} rely on the data-generating process (DGP) being an ARMA process of finite order as long as one assumes stationarity, which we do not have to do with our direct assumption of an AR process.\footnote{Note that different to the notation of \citet{lusompa2023local}, in our case $h=0$ in Equation (\ref{eq: system-of-lps}) defines the one-step-ahead prediction error which is serially uncorrelated by assumption.}
Therefore, robust methods based on either heteroskedasticity and autocorrelation robust (HAC) standard errors \citep[see][]{jorda2005estimation} or adding additional lags of $w_t$ are typically used in practice \citep[see][]{montiel2021local}. Another solution,  proposed by \cite{lusompa2023local}, involves adding estimates of $\{\hat{\varepsilon}_{t+1}, \dots, \hat{\varepsilon}_{t+h-1}\}$ (or using a transformation of the left-hand side variable in the projections) to the $h$-step-ahead regressions to obtain valid inference. 

All these techniques, however, are based on treating each of the forecasting equations as separate estimation problems.\footnote{One exception that takes a Bayesian stance is \cite{tanaka2020bayesian} who relies on a similar representation as we do, and uses it as an estimation device to elicit a joint prior on $\bm b$.} This has implications for the efficiency of the estimators, since $b_i$ and $b_j$ are treated independently of each other for each $i \neq j$. Moreover, impulse responses in dynamic models typically feature patterns such as persistence and smoothness, implying that an independent treatment of each horizon complicates incorporating potentially available prior knowledge about the shape of the IRF.

For these (and other) reasons, our approach is different. Let $\bm y_{t}  = (w_{t}, \dots, w_{t+\tilde{H}})'$; we estimate the system of Equations (\ref{eq: system-of-lps}) jointly and assume a full covariance matrix between the shocks to the different horizons:
\begin{equation*}
    \bm y_{t} = \bm b w_{t-1} + \bm u_{t}
\end{equation*}
with $\bm b = (b_0, \dots, b_{\tilde{H}})'$ and $\bm u_{t} = (u^{(0)}_{t}, \dots, u^{(\tilde{H})}_{t+\tilde{H}})'$. Note that we \emph{jointly} estimate the entire vector of impulse responses $\bm b$, allowing us to easily make probability statements about this relationship between different elements of this vector. Joint inference for impulse responses has received substantial interest --- see \cite{inoue2024inference} for LPs in a frequentist context and \cite{inoue2022} in a Bayesian VAR context, something our approach can naturally deliver. These aspects are discussed in detail in Section \ref{sec:gp-prior}.

For this DGP, introducing a lower triangular $H \times H$ matrix $\bm Q$ given by:
\begin{equation*}
    \tilde{\bm Q} = \begin{pmatrix}
        1 & 0 & \hdots & 0 \\
        \rho & 1 & \hdots & 0 \\
        \vdots & & \ddots & 0\\
        \rho^{h} & \rho^{h-1} & \hdots & 1
    \end{pmatrix},
\end{equation*}
allows us to state the vector of LP residuals $\bm u_{t}$ in terms of the one-step ahead prediction errors of the data-generating process $\bm \varepsilon_{t} = (\varepsilon_t, \dots, \varepsilon_{t+h})'$:
\begin{align}
    \bm u_{t} &= \tilde{\bm Q} \bm \varepsilon_{t},\label{eq: LP_shocks}\\
    \text{Var}(\bm u_t) &= \tilde{\bm Q}~ \underbrace{\text{Var}(\bm \varepsilon_{t})}_{\sigma_\varepsilon^2 \bm I_H}~ \tilde{\bm Q}' = \sigma_\varepsilon^2\tilde{\bm Q}\tilde{\bm Q}'. \label{eq: VC_ut}
\end{align}
While the forecast errors of the AR process are uncorrelated, the residuals of the LPs are correlated across horizons because they represent multi-step forecast errors. Our approach takes into account covariance among the horizon-specific residuals within each stacked LP observation --- Equation (\ref{eq: LP_shocks}) motivates our Bayesian approach to estimate $\bm b$, taking into account the correlation structure between the shocks, determined by $\tilde{\bm Q}$. We achieve this by introducing a \textit{pseudo-likelihood} \citep[see also][who, however, do not jointly estimate the entire system of equations and instead rely on an ex-post adjustment of the posterior variance of impulse responses]{ferreira2023bayesian}, which we build as follows. 

We denote the one-step ahead forecast density function $\hat{p}$, which takes the form:
\begin{equation}
    \hat{p}(\bm y_{t}|\bm b, \bm \Sigma_u) = \mathcal{N}(\bm y_t |  w_{t-1}, \bm b, \bm \Sigma_u) \quad\Leftrightarrow\quad \bm y_{t} = \bm b w_{t-1} + \bm u_{t}, \quad \bm u_t \sim \mathcal{N}(\bm 0_H, \bm \Sigma_u). \label{eq: SUR_rep}
\end{equation}
We can then build the full pseudo-likelihood as
\begin{equation}
\hat{p}(\bm y|\bm b, \bm \Sigma_u)=\prod_{t=1}^T \hat{p}(\bm y_{t}|\bm b, \bm \Sigma_u), \label{eq:SUR_full}
\end{equation}
where we have suppressed dependence on the initial $w_0$ data for convenience.\footnote{Estimates of $\hat{\bm{b}}$ and $\hat{\bm{\Sigma}}_u$ could also be obtained via the generalized method of moments, see, e.g., \citet[][Section 5]{jorda2023local} and \citet{jorda2024local}.} Equation (\ref{eq:SUR_full}) is what we henceforth call a Seemingly Unrelated Local Projection (SU-LP). Two comments are in order: First, the sample size across horizons is not the same because of the different leads on the left-hand side variables. In our estimation approach, we directly tackle this problem and estimate the missing observables to ensure we use all available information. 

Second, at a fundamental level, why is Equation \eqref{eq:SUR_full} a \emph{pseudo-}likelihood and why don't we write down the full likelihood function? The answer is that the set of LPs does not constitute a \emph{generative model} and as such the prediction error decomposition cannot be applied. To see this, note that a standard prediction error decomposition would force us to use $ \mathcal{N}(\bm y_t | \bm  y^{t-1},\bm b, \bm \Sigma_u)$, where the superscript indicates information up to $t-1$. But since $\bm y_t$ and $\bm y_{t-1}$ have elements in common, the prediction error for those elements would be zero, while a standard application of LPs would want to exploit information in the forecast errors for all horizons and time periods. We therefore use the term pseudo-likelihood deliberately. Since the collection of LPs across horizons is not a generative model, there is no full likelihood for this stacked LP object to write down. The pseudo-likelihood is instead a maintained estimation criterion that can discipline estimation of the impulse-response path. We return to this issue and discuss how our approach handles specification issues in more detail in Section \ref{sec: power_posteriors}.

\subsection{Modeling Cross-Horizon Correlations and Dynamic Error Structures}
Writing down the pseudo-likelihood for the entire system {takes into account correlations across horizons}. To see this, it is useful to decompose $\hat{p}(\bm y_{t}|\bm b, \bm \Sigma_u)$ into:\footnote{This decomposition is a prediction-error decomposition, but for one time-period $t$ across all horizons; the problem we just discussed arises when considering a prediction-error decomposition across $t$.}
\begin{equation*}
\mathcal{N}(w_t|w_{t-1},\bm b, \bm \Sigma_u) \cdot \mathcal{N}(w_{t+1}|w_t,w_{t-1},\bm b, \bm \Sigma_u) \cdot \hdots \cdot \mathcal{N}(w_{t+\tilde{H}} |w_{t+\tilde{H}-1}, \hdots, w_t, w_{t-1},\bm b, \bm \Sigma_u).
\end{equation*}
Knowledge of the parameters of the LPs as well as the relevant lags of $w_t$ allows us to explicitly control for correlation across horizons, which we describe in detail below. 

Employing a full error variance $\bm \Sigma_u$ implies that the elements in $\bm u_t$ are correlated and we thus parameterize cross-horizon dependence induced by past shocks.  To see this, we use a decomposition that is related to, but distinct from, that in Equation \eqref{eq: LP_shocks}. In particular, we consider the Cholesky decomposition:
\begin{equation}
    {\bm \Sigma}_u = {\bm Q} {\bm \Omega} {\bm Q}',\label{eq: vardecomp}
\end{equation}
where ${\bm Q}$ is lower triangular with unit diagonal and ${\bm \Omega} = \text{diag}({\omega}_1, \dots, {\omega}_H)$ are the error variances of $\bm e_t$, the uncorrelated errors recovered by the Cholesky decomposition. Note that each horizon has a different variance $\omega_j$ and hence we have heteroskedasticity across the different forecast horizons. This is in contrast to Equation (\ref{eq: VC_ut}). We view this flexibility as an advantage of our approach, even though it is not strictly needed in this specific application, as a comparison of Equations \eqref{eq: LP_shocks} and the expressions we turn to next show. This choice is part of the misspecification discussed above. The unrestricted covariance matrix captures cross-horizon covariance in the stacked system, but it does not impose the exact calendar-time MA structure of overlapping LP residuals.

The Cholesky decomposition enables us to  write the errors in Equation (\ref{eq: SUR_rep}) as:
\begin{equation*}
    \bm u_t = \bm Q \bm e_t, \quad \bm{e}_t \sim \mathcal{N}(\bm 0, \bm \Omega), \quad \bm{e}_t = (e_{t}^{(0)},e_{t+1}^{(1)},\hdots,e_{t+\tilde{H}}^{(\tilde{H})})',
\end{equation*}
and let $q_{ij}$ denote the $(i,j)$th element of the matrix $\bm{Q}$. The first few equations read:
\begin{align*}
    u_t^{(0)} &= e_{t}^{(0)}, \quad u_{t+1}^{(1)} = q_{21} e_{t}^{(0)} + e_{t+1}^{(1)}, \quad u_{t+2}^{(2)} = q_{31} e_{t}^{(0)} + q_{32} e_{t+1}^{(1)} + e_{t+2}^{(2)}, \quad \hdots,\\
    u_{t+h}^{(h)} &= \sum_{i=1}^{h} q_{(h+1)i} e_{t+i-1}^{(i-1)} + e_{t+h}^{(h)}, \quad \text{for } h > 0.
\end{align*}
Hence, the $h$-step-ahead LP forecast errors are a function of the independent errors across the preceding $h-1$ horizons as in Equation (\ref{eq: system-of-lps}). Our pseudo-likelihood controls for $e_t^{(j)}$ for all $j<h$ in the LP for horizon $h$, as these are a function of the parameters of the model and the relevant lags of $w_t$. Throughout, we assume that the horizon $0$ forecast error $u_t^{(0)}$ is serially uncorrelated, a standard assumption in the literature.

To gain a better understanding of the relationship between the different errors, the model can be written in full data notation.  When $e_{t+h}^{(i)} = e_{t+h}^{(j)} = e_{t+h}$, we obtain:\footnote{A variant capable of strictly enforcing this is to use a state space representation and treat the one-step-ahead prediction errors as unobserved states. We provide a brief discussion in Appendix \ref{app:technical}.}
\begin{equation*}
    \underbrace{\begin{bmatrix}
        u_1^{(0)} & u_{2}^{(1)} & \hdots & u_{1+H}^{(H)}\\
        u_2^{(0)} & u_{3}^{(1)} & \hdots & u_{2+H}^{(H)}\\
        & & \vdots & \\
        u_{T}^{(0)} & u_{T+1}^{(1)} & \hdots & u_{T+H}^{(H)}
    \end{bmatrix}}_{\bm{U}} = 
    \begin{bmatrix}
        e_1 & e_2 + q_{21}e_1 & \hdots & e_{1+H} + \sum_{i=1}^H q_{(H+1)i} e_i\\
        e_2 & e_3 + q_{21}e_2 & \hdots & e_{2+H} + \sum_{i=1}^H q_{(H+1)i} e_{i+1}\\
        & & \vdots & \\
        e_T & e_{T+1} + q_{21}e_T & \hdots & e_{T+H} + \sum_{i=1}^H q_{(H+1)i} e_{T+i-1}
    \end{bmatrix},
\end{equation*}
where the $t$th row of $\bm{U}$ is $\bm{u}_t'$. The $j$th (for  $j = h + 1$) column of $\bm{U}$, denoted $\bm{U}_{\bullet j}$, is:
\begin{align*}
    \underbrace{\begin{bmatrix}
        u_{1+h}^{(h)}\\
        u_{2+h}^{(h)}\\
        \vdots\\
        u_{T+h}^{(h)}
    \end{bmatrix}}_{\bm{U}_{\bullet j}} &= 
    \begin{bmatrix}
        q_{j1} & \hdots & q_{jh} & 1 & 0  & \hdots & 0\\
        0 & q_{j1} & \hdots & q_{jh} & 1  & \hdots & 0\\
        \vdots &  & & & & \ddots & \vdots\\
        0 & 0 & \hdots & q_{j1} & \hdots &  q_{jh} & 1\\
    \end{bmatrix}
    \underbrace{\begin{bmatrix}
        e_{1}\\
        e_{2}\\
        \vdots\\
        e_{T+h}
    \end{bmatrix}}_{\bm{e}},
\end{align*}
which demonstrates the MA$(h)$ structure of the reduced form errors at horizon $h$ \citep[see, e.g.,][for a discussion in a Bayesian context]{chan2013moving}. It is worth noting that we use this recursive structure merely for expository purposes. In what follows,  we will consider full-system estimation based on the unrestricted covariance matrix $\bm{\Sigma}_u$, which nests the MA$(h)$ structure but allows for more general correlation structures. 

The AR(1) example abstracts from additional exogenous controls. In practice, researchers often include a large set of additional covariates (which might also include lags thereof) and proxies of economic shocks. In the following sections, we provide computationally tractable methods to carry out estimation with those features and with $\tilde{H}$ taking on large values. 

\subsection{General seemingly unrelated linear projections}
We start by generalizing the ideas laid out in the previous section. Our goal is to estimate the dynamic response of a target variable $w_t$ to a change in a scalar time series $x_t$ conditional on a possibly large panel of additional variables stored in $\bm{r}_t$ and $\bm{s}_t$.\footnote{This framework in principle enables us to compute impulse responses for any number of variables, by replacing $w_t$ in Equation (\ref{eq:components}) with a vector of dependent variables $\bm{w}_t = (w_{1t},\hdots,w_{Mt})'$.} These are of dimension $n_r$ and $n_s$, respectively. The vectors $\bm{r}_t$ and $\bm{s}_t$ are used to distinguish between predetermined and simultaneously determined covariates. Let $\bm{d}_t = (w_t, x_t, \bm{r}_t',\bm{s}_t')'$ and define $\bm{z}_t = (\bm{r}_t',\bm{d}_{t-1}',\hdots,\bm{d}_{t-P}')'$ as a vector of contemporaneous and lagged controls with associated $k\times1$ parameter vector $\bm{\gamma}_{h}$ where $k = n_r + P(2+n_r + n_s)$. 

The general LPs take the form:
\begin{equation}
    w_{t+h} = \beta_h x_t + \bm{\gamma}_{h}'\bm{z}_t + u_{t+h}^{(h)}, \text{ for } h=0, \dots, \tilde{H}.\label{eq:components}
\end{equation}
Note that we have switched the notation relative to the previous section to highlight that it is no longer directly tied to the AR coefficient in the example --- the impulse response of interest is now $\beta_h$, which represents the response of $w_{t+h}$ to an impulse in $x_t$. Although the focus of this paper is on linear LPs, our approach can be easily generalized to allow for nonlinear functions of $x_t$ to enter the LPs as long as $x_t$ is observable (an assumption that we maintain in this section but relax later). Again stacking the horizon-specific LPs yields the general version of SU-LP:
\begin{equation}
    \bm{y}_t = \bm{\beta}x_t + \bm{Z}_t \bm{\gamma} + \bm{u}_t, \quad \bm{u}_t\sim\mathcal{N}\left(\bm{0}_H,\bm{\Sigma}_{u}\right),\label{eq:LP-SUR}
\end{equation}
where $\bm{\beta} = (\beta_0,\beta_1,\hdots,\beta_{\tilde{H}})'$  is our main object of interest, $\bm{Z}_t = (\bm{I}_H \otimes \bm{z}_t')$ is of size $H \times kH$ and  $\bm{\gamma} = (\bm{\gamma}_{0}',\bm{\gamma}_{1}',\hdots,\bm{\gamma}_{\tilde{H}}')'$ includes the coefficients associated with the controls across horizons.

In this general framework, the joint pseudo-likelihood may be obtained as follows. Define $\bm{y} = (\bm{y}_1',\hdots,\bm{y}_{T}')'$, $\bm{X}_t = (\bm{I}_H\otimes x_t)$, $\bm{X} = (\bm{X}_1,\hdots,\bm{X}_T)'$, and $\bm{Z} = (\bm{Z}_1',\hdots,\bm{Z}_T')'$, then we obtain the seemingly unrelated regression (SUR) representation:
\begin{equation}
    \hat{p}(\bm{y}|\bm{\beta},\bm{\gamma},\bm{\Sigma}_u) = \prod_{t=1}^{T}\hat{p}(\bm{y}_t|\bm{\beta},\bm{\gamma},\bm{\Sigma}_u) = \mathcal{N}(\bm{y}|\bm{X}\bm{\beta} + \bm{Z}\bm{\gamma},\bm{I}_T\otimes \bm{\Sigma}_u).\label{eq:SULP-lik}
\end{equation}
We can combine this pseudo-likelihood with suitable priors to derive the respective posterior distributions using Bayes theorem. 

Another, modern interpretation treats the pseudo likelihood as a  Gibbs loss $\ell(\bm \Xi; \bm y) = - \log \hat{p}(\bm y|\bm \Xi)$. \cite{bissiri2016general} show that updating a prior $p(\bm \Xi)$ by this loss function according to:
\begin{equation*}
    p(\bm \Xi|\bm y) \propto \exp(-\ell(\bm \Xi; \bm y)) p(\bm \Xi),
\end{equation*}
represents the solution to a decision problem that trades off expected loss against the Kullback-Leiber distance to the prior (which we define in the next section). Using this interpretation, the posterior distribution of SU-LP can be considered a Gibbs posterior rather than an approximation to an unavailable true posterior distribution.

\subsection{Priors on impulse response functions}\label{sec:gp-prior}
\subsubsection*{General considerations}
SU-LP offers an alternative way of capturing serial correlation in the residuals at the (computational) cost of full-system estimation. Additional potential advantages, such as improvements in efficiency through pooling information across horizons and increased flexibility in terms of prior elicitation also arise. In this section, we will discuss the latter advantage and focus on how the SUR structure can be used to introduce a prior that flexibly controls the shape of the impulse response stored in $\bm \beta$.\footnote{It is important to note that, prior to estimation, we normalize all data to have mean zero and unit standard deviation. This allows us to set priors that are independent of the scale of the observables, making prior elicitation substantially more straightforward. After estimation, we rescale all estimated parameters to take into account the original standard deviations of all variables.}

Typical assumptions underlying stochastic models for dynamic economies imply smooth response functions. Since smoothness of impulse responses is closely linked to dependence between consecutive elements in $\bm \beta$, independence priors of the form $p(\bm \beta) = \prod_{h=0}^{\tilde{H}} p(\beta_h)$ waste known information. Information on the shape of the impulse responses can be introduced explicitly through a general joint prior $ p(\bm \beta)$, which can be decomposed as:
\begin{equation*}
    p(\bm \beta) =  p(\beta_0) \cdot \prod_{h=1}^{\tilde{H}} p(\beta_h|\beta_{h-1}, \dots, \beta_0).
\end{equation*}
The joint prior thus allows for dependence across horizons $h$. Such a joint prior can generally have many parameters because we need to decide how much the impulse responses are correlated across horizons a priori, how much we should shrink towards our prior mean, whether this shrinkage should be horizon-specific, and so on. We now show how to set up a flexible joint prior that only depends on a few hyperparameters that need to be chosen. 

To do so, we use a hierarchical prior. Our joint prior takes the form:
\begin{equation}
    \bm{\beta}|\bm{\mu}_{\beta},\bm{V}_{\beta} \sim \mathcal{N}(\bm{\mu}_{\beta},\bm{V}_{\beta}),\label{eq:irf-prior}
\end{equation}
with prior means $\bm{\mu}_{\beta} = (\mu_{\beta0}, \hdots,\mu_{\beta \tilde{H}})'$ and covariance matrix $\bm{V}_{\beta}$. These prior moments could, in principle, be directly chosen by the researcher as tuning parameters. For instance, setting $\bm \mu_\beta= \bm 0$ and $\bm V_\beta$ such that its Cholesky factor is a lower triangular matrix with ones along and below its main diagonal implies that the prior on $\beta_h$ is centered on $\beta_{h-1}$. Alternatively, $\bm \mu_\beta$ can be set to have a particular shape or centered on the impulse responses of an auxiliary model such as a VAR \citep[see][for a similar approach]{ferreira2023bayesian} or using an empirical Bayes approach centered on the bias-adjusted IRF of \cite{HerbstJohannsen}. In this case, $\bm{V}_{\beta}$ controls the weight put on $\bm{\mu}_{\beta}$ and becomes an important tuning parameter that needs to be chosen manually or by cross-validation.

An alternative route, which we adopt, is to treat both $\bm \mu_\beta$ and $\bm V_\beta$ as unknown and to use hierarchical priors to estimate them. This allows us to remain flexible, yet parsimonious, as we typically only need to choose a small set of prior hyperparameters that parameterize $\bm \mu_\beta$ and $\bm V_\beta$. We introduce a shrinkage prior by assuming that the prior covariance matrix is diagonal, $\bm V_\beta = \diag(v_{\beta0},\hdots,v_{\beta\tilde{H}})$. This allows us to test restrictions of the form $\beta_h = \mu_{\beta h}$ by setting the corresponding element on the diagonal of $\bm V_\beta$, $v_{\beta h}$, appropriately. 

\subsubsection*{Gaussian process priors on impulse response functions}
We assume that $\bm \mu_\beta$ is a smooth function (in a sense we make precise below) and then let the prior variance $\bm V_\beta$ determine the weight to put on $\bm \mu_\beta$. Moreover, given that IRFs might feature certain characteristics (such as being hump-shaped or mean-reverting) we assume that $\bm \mu_\beta$ is a function of the horizons. Specifically, to impose smoothness on our prior impulse responses, we first define a function $\bm \mu_{\beta}(\bm {h})$ that takes a vector $\bm {h}=(0, \dots, \tilde{H})'$ with real, non-negative entries as input. Although this might seem cumbersome at first since we will only use values $h=0,1,...,\tilde{H}$, this assumption allows us to borrow tools from the literature on Gaussian processes (GPs) which are popular in machine learning \citep[for a textbook introduction, see][]{williams2006gaussian}. 

The function $\bm \mu_{\beta}(\bm {h})$ is modeled using GPs:
\begin{equation*}
    \bm \mu_{\beta}(\bm {h}) \sim \mathcal{GP}\left(\underline{\bm \beta}(\bm h), \bm K_{\beta}(\bm h)\right).
\end{equation*}
Here, $\underline{\bm \beta}(\bm h)$ is a mean function, and $\bm K_{\beta}(\bm h)$ is an $H \times H$ kernel. Both typically depend on a small-dimensional vector of hyperparameters, which we do not explicitly indicate as conditioning arguments for notational simplicity. The mean function of the GP can be set in any way one would set $\bm \mu_\beta$ if we chose to treat the prior mean as a deterministic hyperparameter, instead of imposing a hierarchical structure.

Natural choices for $\underline{\bm \beta}(\bm h)$ could be a functional form along the lines of \cite{barnichon2018functional}, possibly informed by previous estimates or economic theory, the prior structure used in \cite{plagborg2019bayesian}, or an empirical Bayes approach where one could center the prior on the de-biased LP estimate of \cite{HerbstJohannsen}, for example. A typical choice, however, is $\underline{\bm \beta}(\bm h) = \bm 0$. This choice is not restrictive since the posterior mean may differ from zero \citep[see][]{williams2006gaussian}, but helpful because we can still impose smoothness via $\bm K_{\beta}(\bm h)$. Thus, we choose $\underline{\bm \beta}(\bm h) = \bm{0}$ as our benchmark, which does not introduce prior information about the location of the IRF. To simplify notation, we omit any dependence on the input vector $\bm{h}$ in what follows.

The GP is a smooth process in $\bm h$ and thus infinite-dimensional. However, given that we only consider a finite number of impulse response horizons, we can rewrite the GP as a multivariate Gaussian prior on $\bm \mu_\beta$: 
\begin{equation}
    \bm \mu_\beta \sim \mathcal{N}(\bm 0, \bm K_\beta).\label{eq:prior_GP}
\end{equation}

Before we discuss the properties of this prior, we state our choice for $\bm K_\beta$. Later we will discuss the distinct roles of $\bm K_\beta$ and the prior variance $\bm{V}_{\beta}$ of the impulse response coefficients conditional on $\bm \mu_\beta$.  Given that we have strong prior views that $\bm \beta$ is smooth, we choose the squared exponential kernel. It has two key hyperparameters. First, the inverse length-scale $\xi > 0$, which is used to set the degree of variability of the prior responses. Setting $\xi$ small implies that the responses are centered on a mean function that is smooth with respect to the forecast horizon whereas larger values of $\xi$ imply that the shape of the mean function displays more variation. The second parameter, labeled $\varsigma \geq 0$ controls how quickly the prior mean response function returns to its unconditional mean.

We define the kernel in two steps, starting with an unscaled version with $(i,j)$th element:
\begin{equation*}
    \tilde{\bm{K}}_{\beta[ij]} = \exp\left(-\frac{\xi \cdot (i - j)^2}{2}\right) \quad \text{for }~i,j = 1,\hdots,H.
\end{equation*}
Hence, depending on $\xi$, we have a specification that implies that the dependence between horizons decreases exponentially. This choice enables us to introduce information on persistence in impulse responses. 

To incorporate prior information on the long-run behavior of the responses, we modify the kernel by introducing additional scaling terms:
\begin{equation}
    {\bm{K}}_{\beta} = \bm{D}^{1/2} \tilde{\bm{K}}_{\beta} \bm{D}^{1/2}, \quad\quad {\bm{K}}_{\beta[ij]} = (d_i d_j)^{\varsigma/2} \cdot \exp\left(-\frac{\xi \cdot (i - j)^2}{2}\right), \label{eq:gp-kernel}
\end{equation}
where $\bm{D}^{1/2} = \diag\left(d_1^{\varsigma/2},\hdots,d_H^{\varsigma/2}\right)$ with $d_i = (H+1-i)/H$ for $i = 1,\hdots,H,$ and we obtain ${\bm{K}}_{\beta[ii]} = d_i^\varsigma$. When $\varsigma = 0$, we have a unit unconditional variance, when $\varsigma = 1$ we have a linearly decreasing variance, whereas larger values of $\varsigma$ exponentially push $\mu_{\beta h}$ towards the unconditional mean of the prior with increasing horizon.

The two hyperparameters $\varsigma$ and $\xi$ play an important role. We illustrate this in more detail in Sub-section \ref{subsec: illustration}. In principle, one can set them so that the estimated IRFs are consistent with some prior view on the shape of the responses. However, given its importance, a natural Bayesian choice would be to elicit yet another set of priors on $\varsigma$ and $\xi$ and sample them alongside the other unknowns of the model. This is what we propose as a baseline. As priors, we use truncated Gaussian priors on both $\xi$ and $\varsigma$:
\begin{align*}
    \xi \sim \mathcal{N}(m_\xi, v_\xi) \cdot \mathbb{I}[\xi \in (\xi_{\text{low}}, \xi_{\text{high}})], \quad 
    \varsigma \sim \mathcal{N}(m_\varsigma, v_\varsigma) \cdot \mathbb{I}[\varsigma \in (\varsigma_{\text{low}}, \varsigma_{\text{high}})],
\end{align*}
where $m_j$ denotes the prior mean and $v_j$ the prior variance for $j \in \{\xi, \varsigma\}$. This prior can be set to be quite uninformative. However, in particular for $\xi$, we found that ruling out too large values improves inference by avoiding cases that imply excessive variation in $\bm \mu_\beta$.

The GP hierarchy should be interpreted as regularization of the impulse-response path, not as a correction that makes the pseudo-likelihood correctly specified. It allows the prior mean to be smooth while permitting horizon-specific departures through the global-local prior. Under standard large-sample conditions, the prior's influence on well-identified finite-dimensional parameters may vanish, so the GP component mainly affects finite-sample regularization. The long-horizon shrinkage governed by $\varsigma$ is therefore an explicit prior choice about the behavior of impulse responses at long horizons.
\subsubsection*{Implications of the kernel}
To see how the kernel in Equation (\ref{eq:gp-kernel}) introduces dependence across horizons, it is worthwhile to move from the \textit{function-space} view of the GP to the \textit{weight-space} view \citep[see][chapter 2]{williams2006gaussian}. First, we can rewrite the prior on $\bm \beta$ as:
\begin{equation}
    \bm{\beta} = \bm{\mu}_{\beta} + \bm{\eta},  \quad \bm{\eta}\sim\mathcal{N}(\bm{0},\bm{V}_{\beta}),\label{eq: prior_beta}
\end{equation}
and then use $\bm \mu_\beta = \bm Q_{\mu_\beta} {\bm \mu}_0$, with $\bm Q_{\mu_\beta}$ denoting the lower Cholesky factor of $\bm K_\beta$ and ${\bm \mu}_0 \sim \mathcal{N}(\bm 0, \bm{I}_H)$, to rewrite Equation (\ref{eq: prior_beta}) as:
\begin{equation}
     \bm{\beta} = \bm Q_{\mu_\beta} {\bm \mu}_0 + \bm{\eta}, \quad \bm \eta \sim \mathcal{N}(\bm 0, \bm V_\beta). \label{eq: weight-space}
\end{equation}
This equation suggests that the prior on $\beta_h$ can be written as:
\begin{equation*}
    \beta_h = \sum_{j=1}^h q_{\mu, hj} ~\mu_{0 j} + \eta_h,
\end{equation*}
where $q_{\mu, hj}$ is the $(h,j)$th element of $\bm Q_{\mu_\beta}$ which is a function of $h$, $j$ and the two hyperparameters $\varsigma, \xi$. Depending on the choice of $\xi$, the lower Cholesky factor could imply that $q_{\mu, h1} < \dots < q_{\mu, h h-1}$ and hence the prior puts less weight on horizon-specific responses that are further in the past.  The presence of the shock $\eta_h$ implies that our setup only pushes the actual response functions towards smooth shapes, but does not strictly impose this.\footnote{Restrictions in the spirit of the distributed lag literature would be imposed such that $\beta_h = \sum_{k=1}^K \tilde{b}_k W_k(h)$ where $W_k(h)$ is a set of $K$ basis functions and $\tilde{b}_k$ are associated weights. Specific choices about $W_k(h)$ combined with a penalized regression approach yield the framework of \citet{barnichon2019impulse}.} Instead, we estimate a (possibly) smooth response function and then shrink $\bm \beta$ toward that function if the data suggests this to be adequate.  

\subsubsection*{Shrinking impulse responses toward  Gaussian processes}
The amount of shrinkage toward the conditional prior mean $\bm \mu_\beta$ is effectively handled through the prior covariance matrix $\bm V_\beta$. 
To see this, notice that Equation (\ref{eq:irf-prior}) can be written as:
\begin{equation*}
    (\beta_h - \mu_{\beta h}) \sim \mathcal{N}(0, v_{\beta h}).
\end{equation*}
If $v_{\beta h}$ is close to zero, $\beta_h$ will be close to $\mu_{\beta h}$ whereas if $v_{\beta h}$ is large, $\beta_h$ is allowed to deviate substantially from $\mu_{\beta h}$ (and the restriction is thus not binding). Hence, setting $v_{\beta h}$ appropriately allows to introduce restrictions on horizon-specific responses. 

We use shrinkage techniques to select $v_{\beta h}$ without requiring much input from the researcher. We do so by using a global-local \citep[GL, see][]{polson2010shrink} shrinkage prior. A GL prior consists of two types of hyperparameters. First, a global shrinkage factor that forces all elements in $\bm \beta$ towards the prior mean $\bm \mu_\beta$. Depending on the parameterization, if this factor is small, without further modifications of the prior, our estimate of $\bm \beta$ would be close to $\bm \mu_\beta$. However, it could be that the horizon-specific estimates $\beta_h$ depart more from $\mu_{\beta h}$, and in this case, using only a global shrinkage parameter would be inappropriate. Hence, we introduce a second type of hyperparameter local to a particular horizon. These pull the estimates away from the prior mean, if necessary.

More formally, we specify the prior variance $\bm V_\beta$ under our GL prior as follows:
\begin{equation}
   \bm V_\beta = \tau^2\cdot\diag(\lambda_0^2,\hdots,\lambda_{\tilde{H}}^2), \quad \text{i.e.,} \quad v_{\beta h} = \tau^2 \lambda_h^2, \quad \text{for }~h=0, \dots, \tilde{H}.\label{eq:prior_Vbeta}
\end{equation}
Here, $\tau$ shrinks globally towards the prior means, whereas local adjustments for horizons are possible through the presence of the $\lambda_h$'s. By specifying suitable priors on $\tau$ and $\lambda_h$ we end up with several popular shrinkage priors used in the machine learning literature. 

We focus on the Normal-Gamma \citep[NG,][]{griffin2010inference} prior. The NG hierarchy is given by:
\begin{equation}
    \tau^2 = 2\tilde{\tau}^{-2}, \quad \tilde{\tau}^2 \sim \mathcal{G}(a_\tau, b_\tau), \quad \lambda_h^2 \sim \mathcal{G}(\vartheta_\lambda, \vartheta_\lambda).\label{eq:prior_NG}
\end{equation}
Here,  $a_\tau, b_\tau > 0 $ and $\vartheta_\lambda > 0$ are hyperparameters chosen by the researcher. Notice that if $\tau^2$ ($\tilde{\tau}^2$) is close to zero (very large), the prior induces much more overall shrinkage. This behavior is obtained by setting $a_\tau$ and $b_\tau$ to small values (a standard choice is, e.g., $a_\tau = b_\tau=0.01$). The parameter $\vartheta_\lambda$ controls the tail behavior of the marginal prior obtained by integrating out the local scaling terms $\lambda_h$. If $\vartheta_\lambda$ is close to zero, the prior puts more mass on zero but the tails become heavier. This implies that in the presence of strong global shrinkage (i.e., $\tau^2 \approx 0$) we still allow for substantial deviations from the prior. If we set $\vartheta_\lambda = 1$ we end up with the Bayesian LASSO \citep[see][]{park2008bayesian}.

\subsection{Illustration of the priors} \label{subsec: illustration}
\begin{figure}[!t]
    \begin{subfigure}[t]{0.49\textwidth}
    \caption{$\varsigma = 0$, $\xi = 0.05$}
    \includegraphics[width=\textwidth]{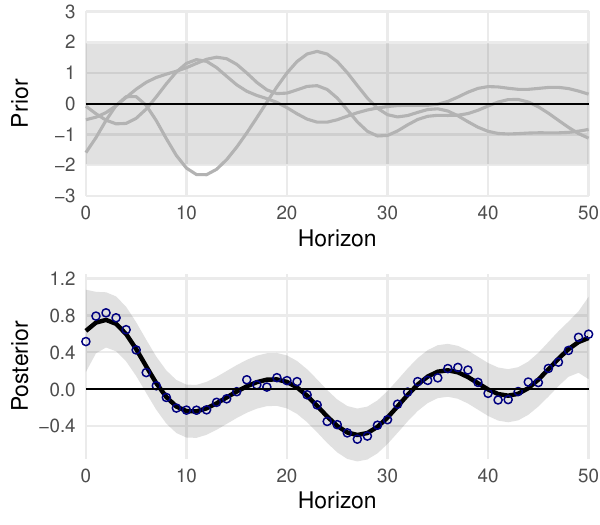}
    \end{subfigure}
    \begin{subfigure}[t]{0.49\textwidth}
    \caption{$\varsigma = 2$, $\xi = 0.05$}
    \includegraphics[width=\textwidth]{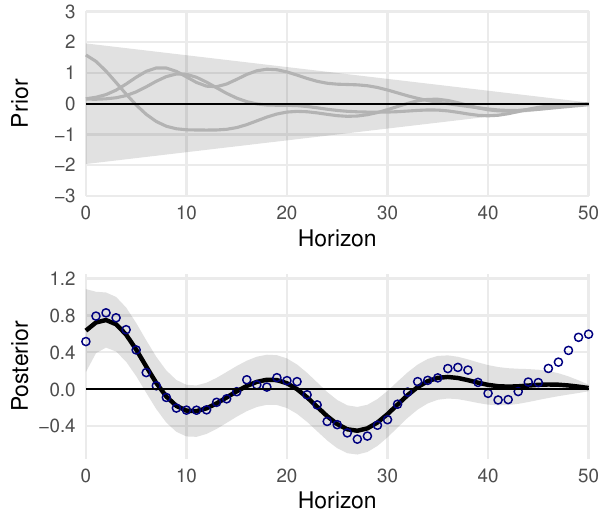}
    \end{subfigure}
    \begin{subfigure}[t]{0.49\textwidth}
    \caption{$\varsigma = 0$, $\xi = 0.005$}
    \includegraphics[width=\textwidth]{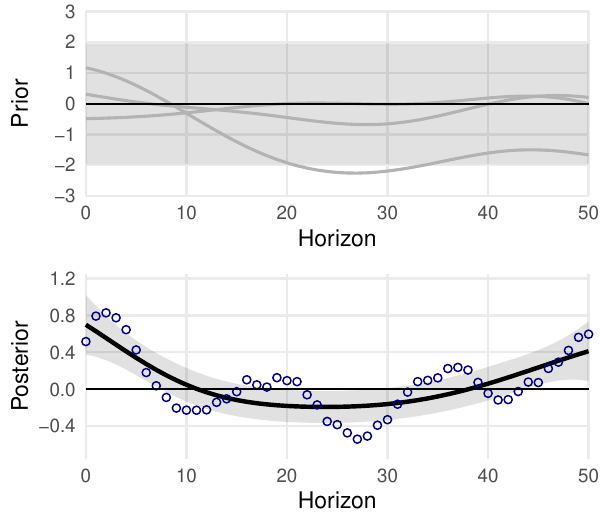}
    \end{subfigure}
    \begin{subfigure}[t]{0.49\textwidth}
    \caption{$\varsigma = 5$, $\xi = 0.05$}
    \includegraphics[width=\textwidth]{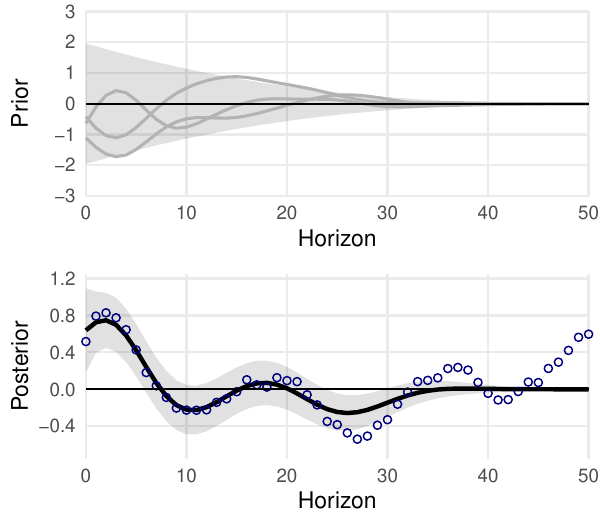}
    \end{subfigure}
    \caption{Gaussian process prior and posterior distribution for various choices of $\varsigma$ and $\xi$ in panels (a) to (d). The posterior is fitted to a dynamic multiplier (blue circles). The gray shaded areas mark the 95\% quantiles, darker gray lines are random draws from the prior while the solid black line indicates the posterior median.}
    \label{fig:gp-example}
\end{figure}

To provide more intuition on how our GP prior works in practice, Figure \ref{fig:gp-example} gives a few examples for different values of $\varsigma$ and $\xi$. The blue circles represent discrete observations of a dynamic multiplier $\bm \beta$, which we use to fit the GPs. Panels (a) to (d) always show the 95\% credible intervals (gray shaded areas) of the prior (upper plot by panel) and the posterior (bottom plot by panel). The gray solid lines refer to three random samples from the prior, whereas for the posterior charts, we indicate the posterior median with a solid black line.

Starting with a comparison of the top panels of Figure \ref{fig:gp-example} (a) and (b) reveals that if we set $\varsigma=0$, the prior variance does not decline with the forecast horizon whereas for $\varsigma = 2$ we shrink the prior credible intervals with $h$, forcing higher-order responses towards zero (or, in general, a pre-specified prior mean). This effect is also visible under the posterior distribution (shown in the bottom panel). If $\varsigma=0$, we observe that the IRFs do not peter out and match $\bm \beta$ which is used as input to train the GP. When we use the prior to force higher-order responses to zero, the posterior is tightly centered around zero and, at least for $h \ge 40$, includes zero.

The effect of $\xi$ on the prior and posterior is best understood by comparing panels (a) and (c).  When $\xi=0.05$, we observe that the dynamic multipliers generated by the prior display more variation. In particular, $\xi$ has an impact on the length of the cycle, with larger values generating shorter cycles. This case is also consistent with the shape of the true responses, translating into a GP posterior estimate that successfully matches the features of the true responses.

When $\xi$ is set to lower values, e.g., $\xi = 0.005$, see panel (c), we find prior responses that are much smoother and display almost no high-frequency variation. This also carries over to the posterior estimates which capture the trend in the true IRFs. The final panel (d) considers the case where $\varsigma$ is set to a large value and therefore the responses for larger horizons are more strongly forced to zero while $\xi$ is set equal to $0.05$.  The prior places substantial mass on zero from horizon $25$ onwards. This also translates into posterior estimates that are tightly centered around zero. This brief discussion shows that the choice of $\varsigma$ and $\xi$ is crucial. 

\begin{figure}[t]
    \includegraphics[width = \textwidth]{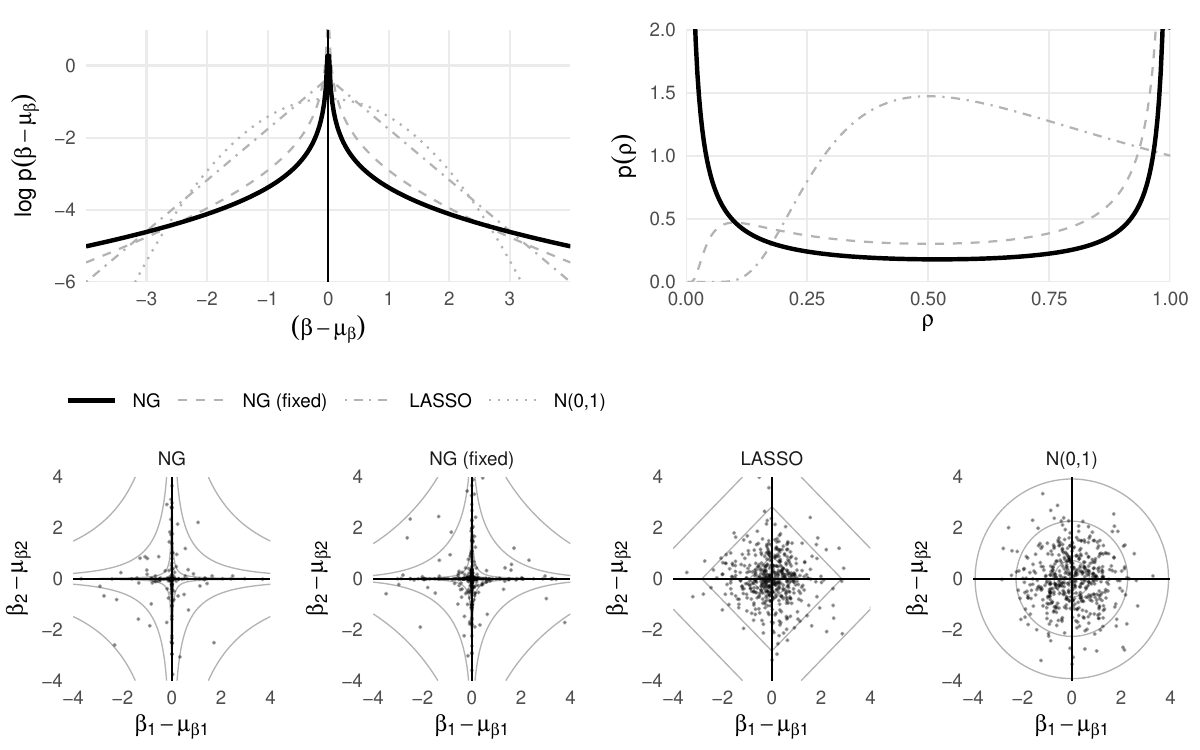}
    \caption{Variants of the global-local prior and implications for shrinkage. The top panels show the log of the marginal priors and the corresponding densities $p(\rho)$ of the shrinkage coefficients $\rho = 1 / (1 + v_{\beta})$. The lower panels show samples and a few contour lines from the bivariate prior $p(\beta_1 - \mu_{\beta 1},\beta_2 - \mu_{\beta 2})$. Prior variants: {{NG} with $\theta_\lambda = 0.1$, $\tilde{\tau}^2 \sim \mathcal{G}(0.1,0.1)$ in solid black; {NG (fixed)} with $\vartheta_\lambda = 0.1$, $\tilde{\tau}^2 = 2$ in dashed gray; {LASSO} with $\theta_\lambda = 1$, $\tilde{\tau}^2 = 2$ in dot-dashed gray; $\mathcal{N}$(0,1) indicates a standard normal prior in dotted gray.} }\label{fig:gl-example}
\end{figure}

Next, we illustrate the shrinkage properties of the prior we use to force $\bm \beta$ towards $\bm \mu_\beta$. We summarize the implications the NG prior has on the difference $\bm \beta - \bm \mu_\beta$ in Figure \ref{fig:gl-example}. To understand differences to other, common, choices such as the LASSO (as a special case of the NG prior) or a Gaussian distributed prior with variance $1$ we also add the corresponding prior shrinkage implications to the figure.

Starting in the top left panel of Figure \ref{fig:gl-example}, we show the log density of the marginal prior obtained after integrating out the local scaling parameters. The log densities indicate that the NG prior (which fixes $\vartheta_\lambda=0.1$ but estimates a global shrinkage factor $\tilde{\tau}^2$) puts substantial mass on zero but, at the same time, allows for large deviations through the heavy tails induced by the prior. This is in contrast with the Bayesian LASSO. In this case, the prior again shrinks most deviations to zero but the thin tails make large deviations highly unlikely. This also shows why the LASSO is theoretically unappealing. It tends to overshrink significant coefficients whereas in the case of small coefficients, it provides too little shrinkage 

Considering the case of a fixed global shrinkage parameter, labeled NG (fixed), reveals a similar shape to the standard NG prior but slightly lighter tails and less shrinkage around the origin. If the prior is standard normally distributed almost no shrinkage is introduced on $\bm \beta - \bm \mu_\beta$. This discussion shows that the NG prior is capable of shrinking deviations strongly to zero if supported by the data. However, if the data suggests substantial deviations of $\beta_h$ from $\mu_{\beta h}$, the prior allows for this through its heavy tails. This is in contrast to priors that induce lighter tails which attribute an extremely low probability to such outcomes.

Next, we consider the shrinkage coefficient $\rho_h = 1/(1+v_{\beta h})$. For illustrative purposes, we consider the simplified case $h=\tilde{H}=0$, and assume that $x_t,u_{t}^{(0)}\sim\mathcal{N}(0,1)$ without additional control variables.\footnote{The general expression is complicated by the presence of covariances across horizons and the variances of the shocks. We provide a discussion of this case in Appendix \ref{app:technical}; see also \citet[][]{polson2010shrink}.} The posterior mean $\overline{\beta}_0$ under these assumptions can be written as $\overline{\beta}_0 = \rho_0 \mu_{\beta 0} + (1 - \rho_0) \hat{\beta}_0$, that is, as a weighted average of the information represented in the data (where $\hat{\beta}_0$ is the least squares estimate) and the prior mean arising from the GP specification. If $v_{\beta 0} \to 0$, then $\rho_0 \to 1$ and we thus end up with $\overline{\beta}_0 = \mu_{\beta 0}$. By contrast, if $v_{\beta 0} \to \infty$ so that $\rho_0 \to 0$ we end up with setting $\overline{\beta}_0 = \hat{\beta}_0$.

With this in mind, consider the different shrinkage profiles induced by the various priors (except for the standard normal prior, which has a fixed variance and is thus excluded from this chart). The profile of the NG prior, which resembles the density of a Beta$(1/c, 1/c)$ distribution (for $c$ being a large number), suggests cases that are either characterized by strong shrinkage so that $\beta_h \approx \mu_{\beta h}$ (corresponding to the case $\rho_0 \approx 1$) or little shrinkage so that $\beta_h$ might deviate strongly from $\mu_{\beta h}$ (so that $\rho_0 \approx 0)$. When we consider the LASSO the pole on $0$ vanishes and we end up with cases in between. This indicates that the LASSO tends to overshrink significant effects (or in our framework to force $\beta_h$ towards $\mu_{\beta h}$ even if not supported by the data) whereas to induce too little shrinkage on cases where the data is consistent with the GP-induced prior. When we fix the $\tilde{\tau}^2$ we end up with a model that either implies very little shrinkage with a mode around $\rho_0 = 0.1$ or a lot of shrinkage (with a pole at $\rho_0 =1$).

The lower panel of Figure \ref{fig:gl-example} is another illustration of the prior and how shrinkage is introduced on different elements of $\bm \beta_h - \bm \mu_\beta$. Each of the panels shows the bivariate prior $p(\beta_1 - \mu_{\beta 1}, \beta_2 - \mu_{\beta 2})$ in the form of a scatter plot and a few contour lines. What these plots reveal is that both NG priors induce much more shrinkage towards the GP but also allow for large deviations, if necessary. This does not carry over to the LASSO and the standard normally distributed prior which either shrink too aggressively, implying many small deviations, or induce no shrinkage at all.

\subsection{Priors on other model parameters}
We assume an inverse Wishart prior on the covariance matrix of the residuals in the LPs:
\begin{equation*}
    \bm{\Sigma}_u \sim \mathcal{W}^{-1}\left(s_0,\bm{S}_0\right),
\end{equation*}
where we set $s_0 = H + 2$ and $\bm{S}_0 = \mathfrak{s}^2(s_0 - H - 1)^{-1}\bm{I}_H$ to guarantee the existence of its moments and $\mathfrak{s}^2 > 0$ is a tuning parameter. On the parameters associated with the control variables we use a conjugate prior setup:
\begin{equation*}
    \bm{\gamma}|\bm{\Sigma}_u \sim \mathcal{N}(\bm{\mu}_{\gamma}, \bm{\Sigma}_u \otimes \bm{V}_\gamma),
\end{equation*}
where $\bm{V}_\gamma$ is a diagonal $k\times k$ prior covariance matrix with known entries. Conjugacy allows for pre-computing several objects required for sampling from the corresponding posterior, which offers significant computational advantages.

How we set the prior moments, $\bm{\mu}_{\gamma} = \text{vec}(\bm{M}_\gamma)$ where $\bm{M}_\gamma$ is of size $k\times H$, and $\bm{V}_\gamma$, is inspired by the Minnesota-tradition. Specifically, we set the prior means to $1$ for the first own lag of $w_t$ (if the variable is in levels, in case it is in differences we set it to $0$), and all remaining elements are zeroes. The informativeness with respect to the controls is set such that we have distinct hyperparameters associated with own and other lags as well as any deterministic variables; in addition, the prior tightness increases with the lag order.

\begin{table}[t!]
    \caption{Summary of the hyperparameters and default choices.}\label{tab:hyperparameters}
    \centering
    \renewcommand{\arraystretch}{1.2}
    \scalebox{0.7}{
    \begin{tabular}{lp{9cm}p{7cm}}
        \toprule
        \textbf{Parameter} & \textbf{Description} & \textbf{Recommendation} \\
        \midrule
        \multicolumn{3}{l}{\textbf{Hyperparameters determining $\bm{K}_\beta$}} \\
        \midrule
        $\xi$ & Length-scale parameter that controls the variability of the GP & Estimate using a truncated normal prior with bounds $(\xi_{\text{low}}, \xi_{\text{high}})$ or fix it to achieve a certain degree of variation in the IRFs \\
        $\varsigma$ & Decay parameter that controls how fast the GP estimate approaches zero & Estimate using a truncated normal prior with bounds $(\zeta_{\text{low}}, \zeta_{\text{high}})$  \\ 
        \midrule
        \multicolumn{3}{l}{\textbf{Prior hyperparameters in case $\xi$ and $\varsigma$ are being sampled}} \\
        \midrule
        $\xi_{\text{low}}$ and $\xi_{\text{high}}$ & Lower and upper bound for the prior on $\xi$ & $\xi_{\text{low}} = 0.01$ and $\xi_{\text{high}}=1$ \\
            $\varsigma_{\text{low}}$ and $\varsigma_{\text{high}}$ & Lower and upper bound for the prior on $\varsigma$ & $\varsigma_{\text{low}} = 0$ and $\varsigma_{\text{high}}=10$ \\
              $m_\xi$ and $v_\xi$ & Prior mean and variance for $\xi$ & $m_\xi= 0.1$ and $v_\xi= 0.1$ \\
            $m_\varsigma$ and $v_\varsigma$ & Prior mean and variance for $\varsigma$ & $m_\varsigma= 0$ and $v_\varsigma= 3$ \\
        \midrule        
        \multicolumn{3}{l}{\textbf{Hyperparameters for $\bm V_\beta$}} \\
        \midrule
        $a_\tau$ and $b_\tau$ & Parameters controlling the overall degree of shrinkage of the Normal-Gamma prior & Set $a_\tau = b_\tau = 0.01$ for heavy shrinkage \\
        $\vartheta_\lambda$ & Parameter controlling the tail behaviour of the prior & Set $\vartheta_\lambda = 0.1$ to induce heavy tails in light of strong shrinkage \\
        \midrule
        \multicolumn{3}{l}{\textbf{Hyperparameters for $\bm \Sigma_u$}} \\
        \midrule
        $s_0$ & Prior degrees of freedom & $H+2$ to ensure a proper prior \\
        $S_0$ & Prior scale matrix & $S_0 = \mathfrak{s}^2 (s_0 - H - 1)^{-1} I_H$ \\
        \midrule
        \multicolumn{3}{l}{\textbf{Hyperparameters for $\bm \gamma$}} \\
        \midrule
        $\mu_\gamma$ & Prior mean on the coefficients associated with the controls & If the target is non-stationary and the first lag of the response is included, set it equal to 1. Otherwise, set everything equal to 0 \\
        $V_\gamma$ & Prior variances for the parameters associated with the controls & Set to resemble features of the asymmetric Minnesota prior \citep[see, e.g.,][]{chan2022asymmetric} \\
        \bottomrule
    \end{tabular}}
\end{table}

Table \ref{tab:hyperparameters} provides an overview of key tuning and hyperparameters. The joint posterior distribution of our framework is not available in closed form, which is why we use Markov chain Monte Carlo (MCMC) methods to sample from it. Most conditional distributions take a well-known form and are thus amenable to Gibbs sampling. Posteriors of parameters that do not follow any easy-to-sample from distributions are updated using Metropolis-Hastings updates. We provide details on posteriors and our full sampling algorithm in Section \ref{sec: sampler} and Appendix \ref{app:technical}.

\section{Making The Model More General}\label{sec: general}
In empirical macroeconomics, researchers are usually faced with situations that depart from the environment described in the previous section. For instance, we implicitly assumed up to this point that the shock series $x_t$ is observed. Unfortunately, this is typically not the case in practice. As a solution, instruments that are correlated with the shock of interest are often available. These are, however, subject to potential measurement errors. In addition, multiple instruments for a single shock of interest may be available. Moreover, researchers might be interested in considering multiple different shocks jointly (so that $\bm x_t$ is a vector). SU-LP is capable of handling these (and more issues) through a few simple modifications on which we focus in this section.

\subsection{Instrumenting shocks}
In practice one often only has access to an instrument $m_t$ that is correlated with the true shocks $x_t$, subject to measurement errors. Our framework can be extended to extract an estimate of the shock of interest based on an instrument by setting up a linear Gaussian state space model.

Let $m_t$ denote an instrument for $x_t$. For the instrument, we invoke the standard relevance and exogeneity conditions \citep{SW_EJ}. Equation (\ref{eq:LP-SUR}) is then complemented by a measurement equation that links $m_t$ to $x_t$:
\begin{equation}
    m_t = \phi x_t + \bm z_t' \bm \delta + \nu_{t}, \label{eq: instrument_1}
\end{equation}
where $x_t \sim \mathcal{N}(0, 1)$ is the (unobserved) structural shock of interest, $\phi$ a coefficient that links the shock to the instrument and $\bm \delta$ is a vector of coefficients associated with the controls in $\bm z_t$ and $\nu_t$ is a white noise shock term with variance $\sigma_\nu^2$. We place an informative inverse Gamma prior on the measurement error variance, $\sigma_\nu^2\sim\mathcal{G}^{-1}(a_{\sigma\nu}, b_{\sigma\nu})$. Such measurement equations appear, for example, in the VAR literature in \cite{Mertens2013,caldara2019monetary,Arias2021}. In contrast to the VAR literature, our approach does not assume invertibility, i.e., we do not assume that structural shocks are a linear function of forecast errors.

In this case, the vector of controls that determines $\bm{z}_t$ is given by $\bm{d}_t = (w_t, m_t, \bm{r}_t',\bm{s}_t')'$, i.e., we include lags of the instrument and not the shock. Notice that under the standard IV assumptions, the marginal variance of $m_t$ can be written as:
\begin{equation*}
    \text{Var}(m_t|\bm z_t) = \phi^2 + \sigma_\nu^2
\end{equation*}
so that the relevance statistic is given by $\phi^2/(\phi^2 + \sigma_\nu^2)$ and thus shows that, for a given $\sigma_\nu^2$, the strength of the instrument increases with $\phi^2$. Stacking Equations (\ref{eq: instrument_1}) and (\ref{eq:LP-SUR}), omitting the control variables for simplicity, yields the state space representation of SU-LP:
\begin{equation*}
    \begin{pmatrix}
        m_t \\
        \bm y_t
    \end{pmatrix} = \begin{pmatrix}
        \phi \\
        \bm \beta
    \end{pmatrix} x_t + \begin{pmatrix}
        \nu_t \\
       \bm u_t 
    \end{pmatrix}, \quad \begin{pmatrix}
        \nu_t \\
        \bm u_t
    \end{pmatrix}
    \sim \mathcal{N}\left(\bm 0, \begin{bmatrix}
         \sigma_\nu^2 & \bm 0' \\
         \bm 0 & \bm \Sigma_u
    \end{bmatrix} \right),
\end{equation*}
so that the shocks $v_t$ and $\bm u_t$ are uncorrelated. This is a (relatively) standard static factor model with a single factor $x_t$. The assumption that the shocks feature unit variance ensures that the scale of $x_t$ is identified. However, the sign of $x_t$ is not identified. We fix the sign of $x_t$ by assuming $\phi > 0$.  It is worth stressing that the structure of SU-LP allows us to estimate $x_t$ alongside the remaining model parameters using straightforward techniques commonly used in the analysis of linear Gaussian state space models \citep[see, e.g.,][]{carter1994gibbs, fruhwirth1994data}. Given that the true shock is a white noise process, obtaining the posterior distribution of $\{x_t\}_{t=1}^T$ is easy and amounts to sampling from a $T$-dimensional Gaussian distribution.

\subsection{Heteroskedastic shocks}
Shocks are most often assumed to be homoskedastic (and typically normalized to have unit variance). Our approach can be straightforwardly extended to allow for heteroskedastic structural shocks. This is done by assuming that:
\begin{equation*}
    x_t \sim \mathcal{N}(0, \sigma_{x,t}^2),
\end{equation*}
where $\log \sigma_{x,t}^2$ is a time-varying (log) volatility factor that evolves according to a standard stochastic volatility (SV) process \citep[see][]{kim1998stochastic, jacquier2002bayesian}:
\begin{equation*}
  \log (\sigma_{x,t}^2) = \rho_x \log (\sigma_{x,t-1}^2) + u_{x,t}, \quad u_{x, t} \sim \mathcal{N}(0, \varsigma^2).
\end{equation*}
Here we let $\rho_x$ denote the persistence parameter and $\varsigma^2$ the variance of the innovations. To fix the scale of $x_t$  in the presence of SV, we assume that the unconditional mean of this process is equal to zero. In this case, the relevance statistic is time-varying and given by:
\begin{equation*}
    \frac{\phi^2 \sigma_{x,t}^2}{\phi^2 \sigma_{x,t}^2 + \sigma_\nu^2}.
\end{equation*}

\subsection{Multiple shocks, their instruments, and measurement errors}
In many cases, interest centers not only on one shock but on multiple shocks jointly. The whole discussion up to this point has been focused on the case of a single shock (or a single instrument per shock). Extending Equation (\ref{eq:LP-SUR}) to allow for multiple shocks and several instruments per shock is straightforward. Suppose that we are interested in estimating the dynamic reactions of $\bm y_t$ to $n_x$ shocks, which we store in $\bm x_t = (x_{1t}, \dots, x_{n_{x}t})'$. Furthermore, suppose that for each shock we have $n_m$ instruments available. Each of these instruments are stored in a $n_m$ vector $\bm m_{it} =(m_{i1,t}, \dots, m_{i n_m, t})'$.\footnote{For simplicity, we assume that each shock has the same number of instruments. In practice, the number of instruments across shocks can, of course, differ, which can easily be incorporated into our framework.} In this case, the equation that links instruments to shocks is given by:
\begin{equation}
    \bm m_{it} = \bm \phi_i x_{it} + \bm z_t' \bm \delta_i + \bm \nu_{it}, \label{eq: IV_LP}
 \end{equation}
with $\bm \phi_i$ denoting a $n_m \times 1$ vector while the remaining terms are defined as in Equation (\ref{eq: instrument_1}) (but per instrument). This observation equation assumes that shock $i$ can be obtained by estimating a single latent factor from a set of competing instruments (that all fulfill the IV assumptions). The resulting factor $x_{it}$ can then be interpreted as an estimate of the shock of interest arising from observing multiple instruments. For such a model to be identified we have to assume the elements in $\bm \nu_{it}$ to be uncorrelated.

The SU-LP representation, in the case of multiple shocks, is given by:
\begin{equation*}
    \bm{y}_t = \sum_{i = 1}^{n_x} \bm{\beta}_i x_{it} + \bm{Z}_t \bm{\gamma} + \bm{u}_t = \bm{X}_t \bm{\beta} + \bm{Z}_t \bm{\gamma} + \bm{u}_t,
\end{equation*}
where $\bm{X}_t = (\bm{I}_H \otimes \bm{x}_t')$ and the shock-specific responses are stored in an $n_x \times H$-matrix $\bm{B} = (\bm{\beta}_1,\hdots,\bm{\beta}_{n_x})'$. The GP priors for the impulse responses, discussed for the single-shock case in Section \ref{sec:gp-prior}, may be defined independently on the $\bm{\beta}_i$'s for $i = 1,\hdots,n_x,$ i.e., the rows of $\bm{B}$. Notice that this general approach also nests the case of a single instrument per shock by setting $n_m = 1$. Moreover, the case that shock $i$ is observed is obtained by setting $n_m = 1$, $\bm \delta_i=0$, $ \bm \phi_i =1$ and Var$(\nu_{it}) =0$. This can be achieved by choosing priors accordingly.

Another issue that commonly arises in applied work is that instruments for different shocks are correlated \citep[see, e.g.,][]{bruns2024avoiding}. To control for this, Equation (\ref{eq: IV_LP}) can be modified as follows. Assuming, for simplicity, that $n_m=1$. Then, we obtain $\bm \nu_t = (\nu_{1t}, \dots, \nu_{n_x t})'$ which follows a Gaussian with zero mean but a (potentially) full covariance matrix $\bm \Sigma_\nu$. In the general case, we assume an informative inverse Wishart prior on $\bm \Sigma_\nu \sim \mathcal{W}^{-1}(s_{0\nu},\bm{S}_{0\nu})$.

Estimating the shocks then boils down to disentangling the uncorrelated components $\bm x_t = (x_{1t}, \dots, x_{n_x t})'$ from a correlated remainder term $\bm \nu_t$. The covariance structure among the instruments under these assumptions is encoded in $\bm \Sigma_\nu$. A key issue is the interpretation of the correlated remainder term $\bm \nu_t$. One possibility is that we assume that the correlation among instruments is entirely due to measurement error. We think this is a particularly appealing assumption when the correlation between instruments is low. On the other hand, we can also exploit comovement among instruments to obtain estimates of common components.

{Indeed, in our empirical work we also use a special case of Equation (\ref{eq: IV_LP}) to extract a common component across shocks. This common component is a model-implied object (such as a common monetary component present in multiple monetary policy instruments or a coordinated monetary/fiscal component when studying monetary and fiscal policy jointly, as we do below). In that case, we propose one of two alternatives, reminiscent of how researchers have incorporated factors in Bayesian time series models: We either explicitly estimate the common factor from $n_m > 1$ instruments by setting $n_x = 1$, or we ex-ante compute the first principal component of all instruments and include that principal component as a joint instrument in our framework. Below we show results for both approaches. An extension to multiple common components is technically also feasible, but would likely lead to weak identification as multiple common components will be hard to disentangle.\footnote{In general the data could be at least somewhat informative about whether or not the common component across instruments is due to noise or a meaningful structural shock as the inclusion of the common component will change the fit of the LP equation.}}

\subsection{Missing data}\label{sec:missing}
So far we have assumed that the dependent variable is observed for all relevant periods and horizons. However, with a fixed sample, longer horizon LPs have to be estimated using fewer observations to account for the lead of the left-hand side variable. We can use the LP structure to efficiently sample unobserved missing values of the dependent variable, circumventing this issue. 

We update the missing leads in the vector of target variables following \citet{chan2023high}, from their joint conditional Gaussian distribution implied by the likelihood defined in Equation (\ref{eq:SULP-lik}). Selection matrices $\bm{\mathrm{S}}_m$ and $\bm{\mathrm{S}}_o$ exist so that $\bm{y}$ can be decomposed into a missing, $\bm{y}_m$, and observed part, $\bm{y}_o$, i.e., such that $\bm{y} = \bm{\mathrm{S}}_m\bm{y}_m + \bm{\mathrm{S}}_o\bm{y}_o$. Moreover, we define $\bm{\beta} = \text{vec}(\bm{B})$ and $\bm{\Sigma}^{-1} = (\bm{I}_T \otimes \bm{\Sigma}_u^{-1})$.

The posterior of the missing values is Gaussian, where $\bullet$ indicates conditioning on all model parameters:
\begin{align}
    \bm{y}_m|\bm{y}_o,\bullet &\sim \mathcal{N}(\bm{\mu}_y,\bm{V}_y),\label{eq:missings_post}\\
    \bm{V}_y &= \left(\bm{\mathrm{S}}_m'{\bm{\Sigma}}^{-1}\bm{\mathrm{S}}_m\right)^{-1}, \quad \bm{\mu}_y = \bm{V}_y\bm{\mathrm{S}}_m'{\bm{\Sigma}}^{-1}\left(\bm{X}\bm{\beta} + \bm{Z}\bm{\gamma} - \bm{\mathrm{S}}_o\bm{y}_o\right).\nonumber
\end{align}
As a by-product, our framework thus yields direct forecasts of the form given in Equation (\ref{eq:components}) for up to $\tilde{H}$-steps ahead, conditional on information at time $T$.

\subsection{Alternative identification schemes}\label{sec:alternativeID}
Up to this point, we focused on the case where the shock is observed or approximated through an instrument. In principle, SU-LP can be used with alternative identification schemes commonly used in the SVAR literature, as other papers in the LP have shown \citep{barnichon2019impulse,plagborg2021local}. A recursive approach is easy to implement by choosing the right control variables in $\bm{r}_t$ and $\bm{s}_t$. Sign restrictions can be imposed by choosing a prior for $\bm{\beta}_i$ that reflects these sign restrictions \citep[for instance, via truncated normal priors, see e.g.][]{baumeister2018inference, korobilis2022new}.

\subsection{Overview Of Our Sampler}\label{sec: sampler}
We now present a stylized representation of our sampling algorithm. Details can be found in Appendix \ref{app:technical}. After initializing all parameters and any latent quantities, our algorithm iterates through the following steps:
\begin{steps}
    \item Updating impulse response and conditional mean parameters
    \begin{itemize}
        \item Sample the impulse response functions (for all shocks of interest, if there are multiple) conditional on all other model parameters, and most importantly on the prior moments driven by the GP, from Equation (\ref{eq:irf_post}). In case we estimate instrument relevance this is included in the block.
        \item Sample the parameters associated with the control variables conditional on everything else from the Gaussian distribution given by Equation (\ref{eq:controls_post}). In case we use an instrumental variable approach, this step includes updating the vector $\bm{\delta}$ in the same block.
    \end{itemize}
    \item Updating covariances and/or variances
    \begin{itemize}
        \item The covariance matrix across horizons is sampled using $\bm{U}$ from Equation (\ref{eq:Sigu_post}).
    \item The variances (or covariance matrix) of the measurement errors, if applicable, can be estimated from their inverse Wishart posterior (in the case of non-zero covariances); or from independent inverse Gamma distributions.
    \end{itemize}
    \item Updating the hierarchical prior (GP and GL shrinkage prior)
    \begin{itemize}
        \item Based on the draws of the impulse response functions at the lowest hierarchy, we may update the GP using Equation (\ref{eq:post_GP}) on a shock-by-shock basis.
        \item The hyperparameters associated with the GP prior are updated using Metropolis-Hastings steps as described in the context of Equation (\ref{eq:post_GPhyperMH}).
        \item The variances determined by the GL prior are sampled from Equation (\ref{eq:post_NG}).
    \end{itemize}
    \item Updating latent state variables
    \begin{itemize}
        \item Conditional on everything else, and if we assume latent shocks, it is straightforward to update the $x_{it}$'s from their Gaussian posterior distribution. In case we assume heteroskedastic errors, we may update the log-volatility processes conditional on the full history of the states.
        \item We may sample the missing values in $\bm{y}$ using the Gaussian distribution in Equation (\ref{eq:missings_post}). Since most of the involved matrices are banded and/or sparse, we update these quantities efficiently using precision sampling.
    \end{itemize}
\end{steps}

In our empirical illustrations, we cycle through our algorithm $12,000$ times and discard the first $3,000$ draws as burn-in. All inference is then based on each $3$rd of the remaining draws. Given the conditionally conjugate structure of our model, mixing for most parameters appears to be no issue in most cases, as measured by inefficiency factors. The only exceptions are the hyperparameters associated with the kernel of the GP. In this case, mixing is slightly worse but still acceptable. For our most general specification with multiple shocks, it takes about $14$ seconds to obtain $1,000$ draws from the posterior on a Macbook Air M1.

\subsection{Pseudo likelihoods, mis-specification and power posteriors} \label{sec: power_posteriors}
Using a pseudo-likelihood $\hat{p}(\bm y|\bm \Xi)$, with $\bm \Xi$ denoting all model parameters (and possibly static factors integrated out analytically), means that posterior inference is based on a misspecified estimation criterion rather than on the likelihood of a generative model for the stacked LP equations.

As discussed in the previous sub-sections, SU-LP offers two mechanisms that aim to make this misspecified criterion useful for estimation. These are the system-based estimation approach that uses cross-horizon covariance information and the non-parametric prior on $\bm \beta$ that provides additional flexibility. We will show in the next section that these features indeed lead to empirical coverage rates that are close to the nominal ones.  

However, if the researcher believes that this is not enough (for instance, if the DGP features structural breaks, non-linear features, non-Gaussian shocks, or omitted variables), one option would be to robustify SU-LP towards different (but unknown) forms of misspecification. A widely used approach in Bayesian statistics and machine learning that can address these challenges is based on the so-called power posterior \citep{holmes2017assigning, grunwald2017inconsistency, medina2022robustness}:
\begin{equation}
    p_c(\bm \Xi \mid \bm y) = \frac{\hat{p}(\bm y \mid \bm \Xi)^c \, p(\bm \Xi)}{p(\bm y \mid c)},\label{eq: power_posterior}
\end{equation}
for a learning rate $c \in (0, 1]$. A value of $0$ implies that no learning takes place (and the power posterior equals the prior) while  $c=1$ gives the standard posterior obtained through the MCMC algorithm outlined in Section \ref{sec: sampler}. Intermediate values imply that the likelihood is downweighted and less weight is put on data-based information. Essentially, smaller values of $c$ protect posterior inference against significant levels of misspecification whereas setting $c$ close to one is appropriate when the pseudo-likelihood is a useful approximation for the target estimand. Sandwich-adjusted posteriors provide a related route for reducing frequentist risk under misspecification \citep{muller2013risk}; here we instead study a power-posterior adjustment.

In simple models, using the power posterior is easy and posterior simulation can be carried out conditional on a particular value of $c$. However, in our hierarchical model that consists of multiple layers (with several latent quantities), we opt for a different approach that takes the standard posterior (i.e., with $c=1$) and ex-post modifies it to end up with draws from the posterior for $c < 1$. Our approach builds on importance sampling and uses the standard posterior as an importance density and  weights that depend on $\hat{p}(\bm y|\bm \Xi)^{c-1}$.\footnote{For a similar approach to setting the learning rate of a power posterior, see \cite{sona2023quantification}.}

These weights are obtained by dividing Equation (\ref{eq: power_posterior}) by $p(\bm \Xi|\bm y) = \hat{p}(\bm y|\bm \Xi) p(\bm \Xi) / p(\bm y)$, noting that the ratio of the power marginal likelihood $p(\bm y|c)$ and the standard marginal likelihood $p(\bm y)$ does not depend on $\bm \Xi$, leads to:
\begin{equation*}
\frac{p_c(\bm \Xi| \bm y)}{p(\bm \Xi|\bm y)} \propto \hat{p}(\bm y | \bm \Xi)^{c -1}.
\end{equation*}
If we'd like to compute $E_c(g(\bm \Xi)) = \int g(\bm \Xi) p_c(\bm \Xi|\bm y) d \bm \Xi$, multiplying by $\hat{p}(\bm \Xi| \bm y) \times 1/\hat{p}(\bm \Xi|\bm y)$ results in:
\begin{align*}
E_c(g(\bm \Xi)) &= \int \frac{g(\bm \Xi) p_c(\bm \Xi|\bm y) p(\bm \Xi|\bm y) }{p(\bm \Xi| \bm y) } d \bm \Xi,  
\end{align*}
which, after using $\hat{p}(\bm y|\bm \Xi)^{c} = \hat{p}(\bm y|\bm \Xi)^{c-1} \times \hat{p}(\bm y| \bm \Xi)$, equals 
\begin{equation}
    E_c(g(\bm \Xi)) = \int g(\bm \Xi) \hat{p}(\bm y|\bm \Xi)^{c-1} p(\bm \Xi|\bm y) d \bm \Xi, \label{eq: IS}
\end{equation}
which is the importance sampling identity with the proposal distribution $p(\bm \Xi| \bm y)$ and $p_c(\bm \Xi| \bm y)$ being the target density.  

Let $\bm \Xi^{(s)}$ denote the $s$th draw from the SU-LP posterior. Given a sequence of $S$ draws, we can approximate (\ref{eq: IS}) as:
\begin{equation*}
\hat{E}_c(g(\bm \Xi)) = \sum_{s=1}^S g(\bm \Xi^{(s)}) ~ w^{(s)},
\end{equation*}
where $w^{(s)} = \hat{p}(\bm y|\bm \Xi=\bm \Xi^{(s)})^{c-1}/\sum_{j=1}^S \hat{p}(\bm y|\bm \Xi=\bm \Xi^{(j)})^{c-1}$.  After obtaining weights, the posterior $p_c(\bm \Xi| \bm y)$ can then be approximated numerically by sampling (with replacement) from $p(\bm \Xi| \bm y)$ with weights $\bm w = (w_1, \dots, w_S)'$. Adding this step involves minimal computational overhead. However, as is common for importance sampling techniques, it could be that the weights become degenerate. To see this, note that for $c<1$, the power posterior is more diffuse than the standard posterior. Using $p(\bm{\Xi}|\bm{y})$ as a proposal therefore means the proposal is more concentrated than the target. In that situation, the importance weights  can have high variance if $c$ is sufficiently below $1$ or if the likelihood is  sharply concentrated, leading to a small effective sample size.  In our simulations, we show that for values of $c > 0.7$, slightly increasing the number of retained draws from the standard posterior yields appropriate effective sample sizes. Since this has to be done once, we consider it relatively unproblematic. 

There are several ways to choose the value of \( c \). Some methods --- such as the SafeBayes approach of \citet{grunwald2017inconsistency} --- require cross-validation, which can make the selection of \( c \) computationally intensive. Fortunately, in our setting, we can evaluate \( p_c(\boldsymbol{\Xi} \mid \boldsymbol{y}) \) easily for a range of values of \( c \), allowing for alternative strategies. One option is to choose \( c \) such that the posterior credible intervals of \( p_c(\boldsymbol{\beta} \mid \boldsymbol{y}) \) match the width of uncertainty bands produced by commonly used robust LP estimators. Alternatively, we may select \( c \) to ensure that the resulting credible intervals are conservative while still yielding statistically significant impulse responses at the horizons of interest. In that case, the chosen value of \( c \) is itself informative, quantifying the amount of evidence required to detect a significant effect. A third approach is to use simulations based on a realistic data-generating process and select \( c \) to minimize the distance between actual and nominal coverage rates.

\section{A Monte Carlo study} \label{sec: monte_carlo}
\subsection{Monte Carlo design}
In this section, we analyze the performance of SU-LP (and the variant based on the power posterior) using a realistic DGP that aims to mimic the dynamics of US macroeconomic data. Following \cite{schorfheide2005var} and \cite{gonzalez2025misspecification}, we assume a VARMA($P,\infty$) process for an $n$-dimensional vector of observables $\bm{w}_t$, for a sample $t = 1,\hdots,T$:
\begin{equation*}
    \bm{w}_t = \sum_{p = 1}^P \bm{\Phi}_p \bm{w}_{t-p} + \bm{H}\bm{\epsilon}_t + \alpha T^{-\pi} \sum_{j = 1}^\infty \bm{A}_j\bm{H}\bm{\epsilon}_{t-j}, \quad \bm{\epsilon}_{t}\overset{\text{iid}}{\sim}\mathcal{N}(\bm{0},\bm{I}_n),
\end{equation*}
with $n\times n$ dynamic coefficients $\bm{\Phi}_p$, $\bm{H}$ is an $n\times n$ structural impact matrix, and $\bm{A}_j$ are $n\times n$ MA coefficients. The weight on the MA part is a function of the sample size, implying that the importance of the MA term declines with increasing sample sizes. The term $\alpha$ measures the overall weight; if the researcher estimates a standard VAR, then $\alpha$ measures the overall degree of misspecification and $\pi$ measures local misspecification.

To closely mimic US macroeconomic dynamics, we partially calibrate the parameters using a reduced form VAR$(P)$. We use $n = 7$ macroeconomic and financial variables transformed towards approximate stationarity for the US economy, choose $P = 5$ lags, and obtain $\bm{H}$ with a Cholesky decomposition of the reduced-form covariance matrix. Subsequently, we use the posterior median estimate of any calibrated parameters, set $\pi = 0.5$, and the MA coefficients are simulated from independent standard normal distributions (up to a maximum order of $10$, for $j > 10$ they are set equal to $0$). 

We rely on $1000$ replications for $T + 1000$ periods and discard the first $1000$ observations, which leads to samples of length $T$. We consider four sample sizes, $T \in \{100,250,500,1000\}$, and two cases for the weight on the MA part, $\alpha \in\{0,2\}$. We compute impulse responses up to a maximum horizon $\tilde{H} = 16$, and refer to the true IRF as $\bm{\beta}_{\ast}$ below. 

Similarly to our applications with real data, we focus on the responses of the output growth variable to a shock in the monetary policy rate. Moreover, we consider the output responses for four models.
The first is the proposed GP and GL prior setup, referred to as SU-LP. The second is an implementation with a ``flat'' prior that sets $\bm{\mu}_\beta = \bm{0}_H$ and $\bm{V}_\beta = 10\bm{I}_H$,  labeled SU-LP (flat). The third is the classical LP estimated with OLS using \cite{NW1987} HAC standard errors (LP, default). The fourth is the regularized classical LP, following \citet{barnichon2019impulse}, with a penalty matrix that shrinks towards a line ($r = 2$ in their notation; labeled LP, smooth). Consistent with the original paper, we select the hyperparameters using cross-validation. The exercise does not include bootstrap LP intervals, HAR standard errors \citep{lazarus2018har}, or sandwich-adjusted Bayesian posteriors \citep{muller2013risk}; these are natural benchmarks for a broader frequentist evaluation, especially in light of small-sample bias and size distortions in LP inference \citep{HerbstJohannsen}.


Additional details such as specific details on the DGP calibration, the computation of true IRFs, and associated charts of example DGP realizations are provided in Appendix \ref{app:technical}. Appendix \ref{app:empirical} includes another small-scale simulation exercise based on a DSGE-based DGP.

\subsection{Monte Carlo results}
Table \ref{tab:coverage} shows the coverage probabilities of the 90 percent posterior credible set or confidence interval across different LP implementations for $1000$ replications of the DGP for selected horizons. To avoid mixing up effects of structural identification, measurement errors, and statistical estimation, we treat the true shock as observed for all competing specifications in this analysis.

The table suggests that for very short samples ($T=100$) and $\alpha \in \{0, 2\}$, both variants of SU-LP produce coverage rates that are much closer to the nominal 90 percent level than the OLS-based LPs with HAC standard errors and the smooth LPs of \cite{barnichon2019impulse}, both of which produce coverage rates substantially below 90 percent. Comparing both SU-LP specifications reveals that the specification with a flat prior on $\bm \beta$ performs slightly better for horizons up to six steps ahead. For larger impulse response horizons, the pattern is mixed and shrinking the IRFs towards a Gaussian process produces slightly more favorable coverage ratios for $h \in \{8, 10, 12\}$ before producing slightly too wide credible intervals for longer-run IRFs.

\begin{table}[t!]
    \includegraphics[width = \textwidth]{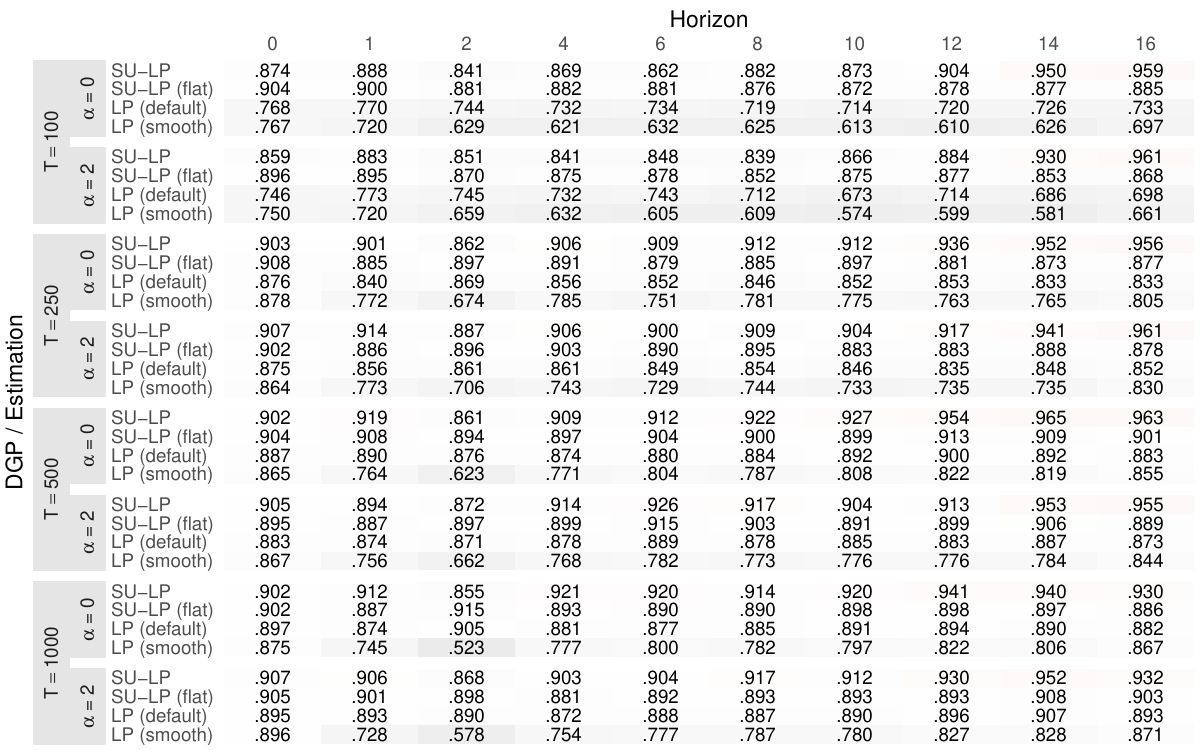}
    \caption{Coverage probabilities of the 90 percent posterior credible set or confidence interval across different LP implementations for $1000$ replications of the DGP for selected horizons. Sample size $T \in \{100,250,500,1000\}$ and weight on MA part $\alpha \in\{0,2\}$. Gray shading indicates undercoverage, red shading vice versa.}
    \label{tab:coverage}
\end{table}

When we increase the length of the sample to $T=250$ and $T=500$, LP (default) produces better calibrated credible intervals with coverage rates  closer to 90 percent.  However, both variants of SU-LP yield favorable coverage rates for (most) forecast horizons considered in these DGPs. Interestingly, we find no discernible differences between DGPs that set $\alpha=0$ or $\alpha=2$.

The comparatively strong performance of SU-LP also carries over to very large samples ($T=1000$). In this case, SU-LP and SU-LP (flat) produce coverage rates close to 90 percent. LP (default) also produces well calibrated intervals. Only LP (smooth) undercovers and produces too narrow credible intervals, a finding that is consistent with simulation results in \cite{barnichon2019impulse}.

\begin{figure}[t!]
    \includegraphics[width = \textwidth]{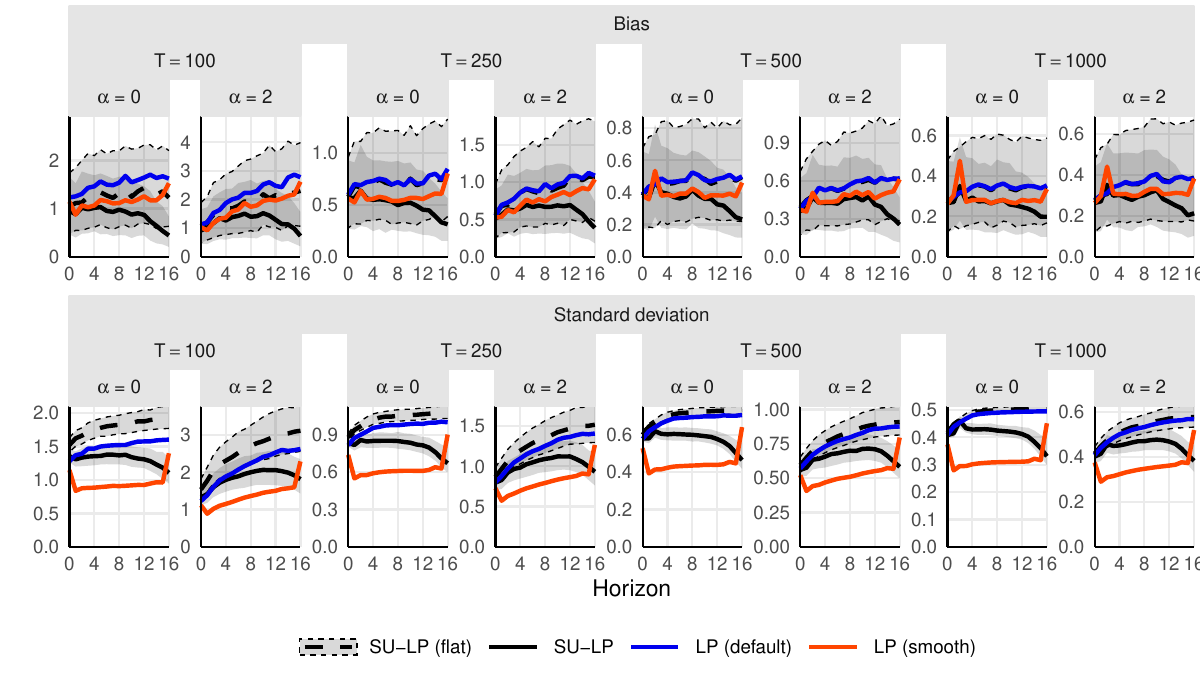}
    \caption{Median (and 25th and 75th percentile for SU-LP variants) absolute bias $\mathbb{E}(\bm{\beta} - \bm{\beta}_{\ast})$ and standard deviation $\text{Var}(\bm{\beta})^{1/2}$ normalized by $(\bm{\beta}_{\ast}'\bm{\beta}_{\ast} / \tilde{H})^{1/2}$ across $1000$ replications of the DGP. Sample size $T \in \{100,250,500,1000\}$ and $\alpha \in\{0,2\}$.}
    \label{fig:normbias}
\end{figure}

Coverage rates tell us whether our model produces credible intervals surrounding the LPs that are well calibrated. They tell us nothing about the error we make when producing point estimates of the IRFs. In principle, a successful approach should produce a low bias under relatively low risk (defined as the standard deviation of the estimator). We investigate this relationship systematically in Figure \ref{fig:normbias}. The figure shows the median (and 25th and 75th percentiles for SU-LP variants) absolute bias, defined as $\mathbb{E}(\bm{\beta} - \bm{\beta}_{\ast})$, and the standard deviation $\text{Var}(\bm{\beta})^{1/2}$ normalized by $(\bm{\beta}_{\ast}'\bm{\beta}_{\ast} / \tilde{H})^{1/2}$, inspired by \citet{li2024local}, for the different LP variants across $1000$ realizations from the DGP. Thus, Figure \ref{fig:normbias} reports bias and dispersion separately, which speaks to the same bias--variance trade-off as MSE, although we do not add a separate MSE table here.

Starting with the bias in the top panel of \autoref{fig:normbias}, there are four notable insights. First, regardless of the value of $\alpha$ and all sample sizes, there are only minor differences in bias with respect to impact estimates. However, when we extend the impulse response horizon, a consistent picture emerges where SU-LP produces the lowest bias for $h \ge 4$. Second, when only bias is considered, the default LP estimator exhibits the weakest performance among all models considered. In all cases and for most horizons, it is the approach that produces the largest bias. Third, comparing SU-LP and SU-LP (flat) shows that forcing the IRF estimator towards a Gaussian process pays off, particularly for longer impulse response horizons. LP (smooth) displays a performance comparable to that of SU-LP (flat). Fourth, relative differences across DGPs are small and we mostly find that if the DGP suggests misspecification, most models produce a larger bias. However, the main exception is SU-LP which, irrespective of the value of $\alpha$, produces biases that are barely distinguishable from each other for different values of $\alpha$ (but conditional on $T$). This corroborates our narrative that adding a GP component to the model alleviates bias arising from misspecification.

\cite{li2024local} state that bias reduction is no free lunch, meaning that methods that produce low bias do so at the cost of more estimation uncertainty. Intuitively speaking, we would expect methods that perform well in terms of bias to be accompanied by larger standard deviations. However, this only holds for SU-LP (flat) and (with an opposite sign) for the standard LP estimator. Particularly for SU-LP, we find that is produces a low bias with the second lowest standard deviation across all DGPs considered and for (most) horizons considered. This, in combination with the coverage rates, paints a favorable picture of the statistical properties of SU-LP which produces accurate IRF estimates with reasonably calibrated credible intervals. 

\subsection{Assessing power posteriors}
Our Monte Carlo results suggest that SU-LP performs well across common types of misspecification considered in the literature. However, there may be other types of misspecification that are even more severe. In this case, the researcher can use the power posterior outlined in Section \ref{sec: power_posteriors}.

In this section, we evaluate how the power posterior performs for various $c$ values. Our emphasis remains on coverage rates, bias, and standard deviations, particularly examining the SU-LP configuration with GP and GL priors. The Monte Carlo analysis indicated negligible (qualitative) differences in SU-LP's performance across different sample sizes, so we use $T=250$, which is similar to the number of observations available in common quarterly macroeconomic datasets.

\begin{figure}[t!]
    \begin{subfigure}[t]{\textwidth}
    \caption{}
    \includegraphics[width=\textwidth]{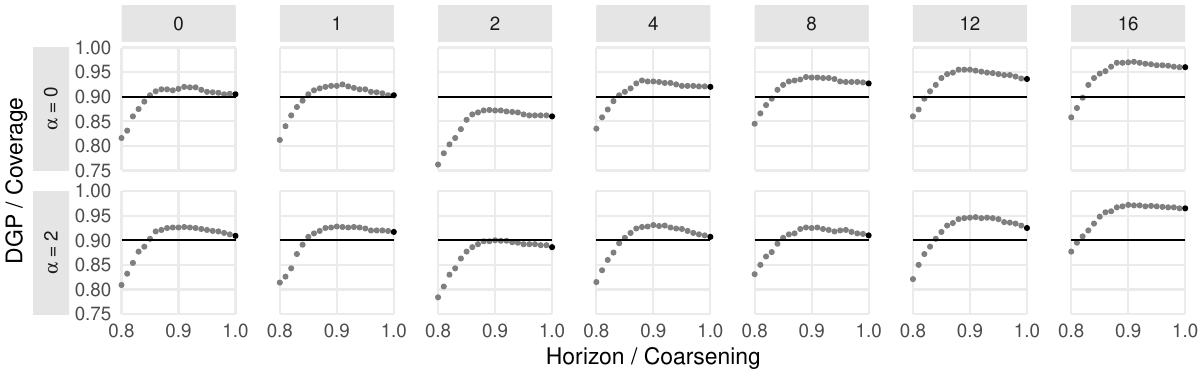}
    \end{subfigure}
    \begin{subfigure}[t]{\textwidth}
    \caption{}
    \includegraphics[width=\textwidth]{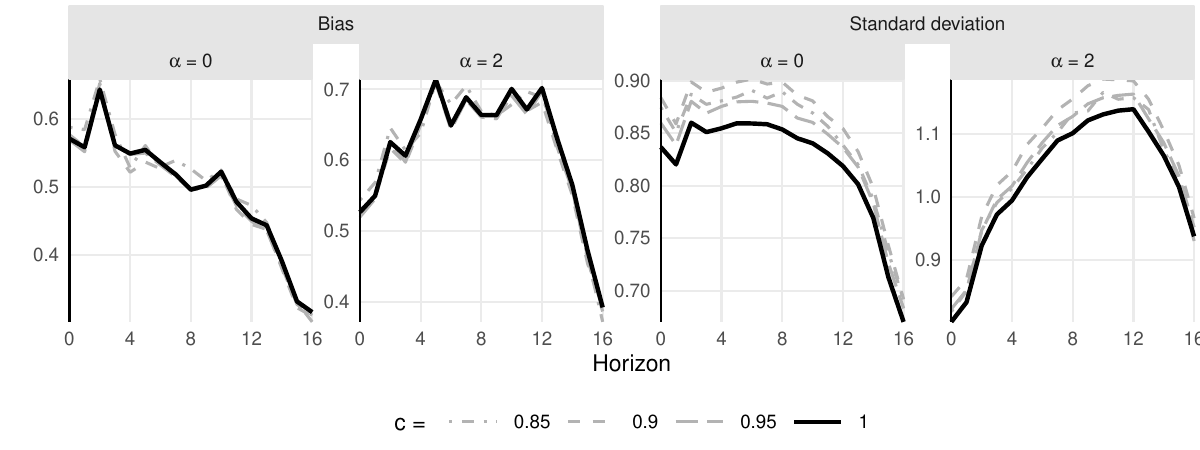}
    \end{subfigure}
    \caption{Coverage probabilities of the 90 percent posterior credible set in panel (a) for different settings of the coarsening parameter $c \in \{0.80,0.81,\hdots,1.00\}$. Median absolute bias $\mathbb{E}(\bm{\beta} - \bm{\beta}_{\ast})$ and standard deviation $\text{Var}(\bm{\beta})^{1/2}$ normalized by $(\bm{\beta}_{\ast}'\bm{\beta}_{\ast} / \tilde{H})^{1/2}$ in panel (b). Results across $1000$ replications with sample size $T  = 250$ and $\alpha \in\{0,2\}$.}
    \label{fig:coarseningsim}
\end{figure}

Our results across $1000$ draws from the DGP and for $\alpha \in\{0,2\}$ are structured similar to the preceding simulation study. Figure \ref{fig:coarseningsim} shows coverage rates of the 90 percent posterior credible set in panel (a) for different settings of the coarsening parameter $c \in \{0.80,0.81,\hdots,1.00\}$. Note that $c = 1$ refers to the standard posterior. In panel (b) we plot the normalized median absolute bias and standard deviation for selected values of $c$.

Panel (a) of \autoref{fig:coarseningsim} suggests that if the DGP features no MA term, using the power posterior hurts in terms of coverage as long as $c < 0.85$. For $c=0.85$, we obtain coverage rates close to 90 percent for most forecast horizons. Notice that for $h \le 4$, the standard posterior produces coverage rates very close to 90 percent as well. Only in the case of longer-run IRFs ($h \ge 8$) we find that the model based on the power posterior produces coverage rates close to the nominal one, whereas $c=1$ produces slightly inflated credible intervals. This story, albeit to a lesser degree, carries over to the DGP that sets $\alpha=2$. In this case, we again find that $c=0.85$ is a good choice in terms of achieving coverage close to $90$ percent. The standard posterior performs well throughout, with coverage rates close to (or slightly) above $90$ percent.

Considering bias and standard deviation (see panel (b)) reveals that the standard posterior is highly competitive, producing a bias close to the one of the power posterior with $c=0.85$ but with substantially smaller standard deviations. This holds for both values of $\alpha$. 

In summary, our simulation results demonstrate that SU-LP maintains robust performance, even in DGPs with MA terms. While introducing the power posterior might offer some improvements, the increase in coverage is generally minor. As for bias and standard deviations, such enhancements are even less significant. Consequently, when using US macroeconomic time series data similar to the one we use to calibrate the DGP, we suggest using the standard posterior.

\section{Empirical Applications}\label{sec: applications}
In all of our empirical work below, we do not consider contemporaneous control variables, i.e., $\bm{r}_t = \emptyset$.  We produce IRFs using the same four LP estimators (SU-LP, SU-LP (flat), LP (default) and LP (smooth)) as in the Monte Carlo exercise.

Throughout we scale responses so that they can be interpreted as the reaction of $\bm{y}_t$ to a one unconditional standard deviation increase in $x_t$. This choice ensures that IRFs are comparable over time if we assume heteroskedastic shocks. Notice that we do not scale the impulse responses back into the scale of $m_t$.

\subsection{Monetary policy in the US}
In our empirical work with real data, the LPs, which Equation (\ref{eq:components}) specifies in levels, are estimated using long differences, see, e.g., \citet{piger2023differences}:
\begin{equation*}
    (w_{t+h} - w_{t-1}) = \beta_h x_t + \bm{\gamma}'(\bm{z}_{t} - \bm{z}_{t-1}) + u_{t+h}^{(h)},
\end{equation*}
which have been shown to yield better small-sample frequentist results when the data is persistent. Using $\Delta\bm{y}_t = [(w_{t},\hdots,w_{t+\tilde{H}}) - (\bm{\iota}_{H}' \otimes w_{t-1})]'$ and $\Delta\bm{Z}_t = [\bm{I}_H \otimes (\bm{z}_{t} - \bm{z}_{t-1})']$, we obtain the corresponding SUR representation:
\begin{equation*}
    \Delta\bm{y}_t = \bm{\beta} x_t + \Delta\bm{Z}_t\bm{\gamma} + \bm{u}_t.
\end{equation*}

For this illustration, we estimate the response of a single target variable, real GDP (FRED acronym  \texttt{GDPC1}, stored in the variable $Y_t$), transformed as $w_t = 100\cdot\log(Y_t)$, to a monetary policy shock. The control variables $\bm{s}_t$ include $P=5$ lags of CPI inflation, the S\&P 500 index (both as $100\cdot\log$), the federal funds rate and BAA spreads alongside lags of the target variable and the instrument. In addition, we add a linear time trend and an intercept. All variables are taken from the \href{https://research.stlouisfed.org/econ/mccracken/fred-databases/}{FRED-QD dataset} and measured on a quarterly frequency ranging from 1970Q1 to 2007Q4. We rely on the well-known \citet{romer2004new} shocks \citep[updated by][]{wieland2020financial} in $m_t$; for the variants of our Bayesian framework, we use latent heteroskedastic shocks and further assume iid measurement errors.

\begin{figure}[!t]
    \includegraphics[width=\textwidth]{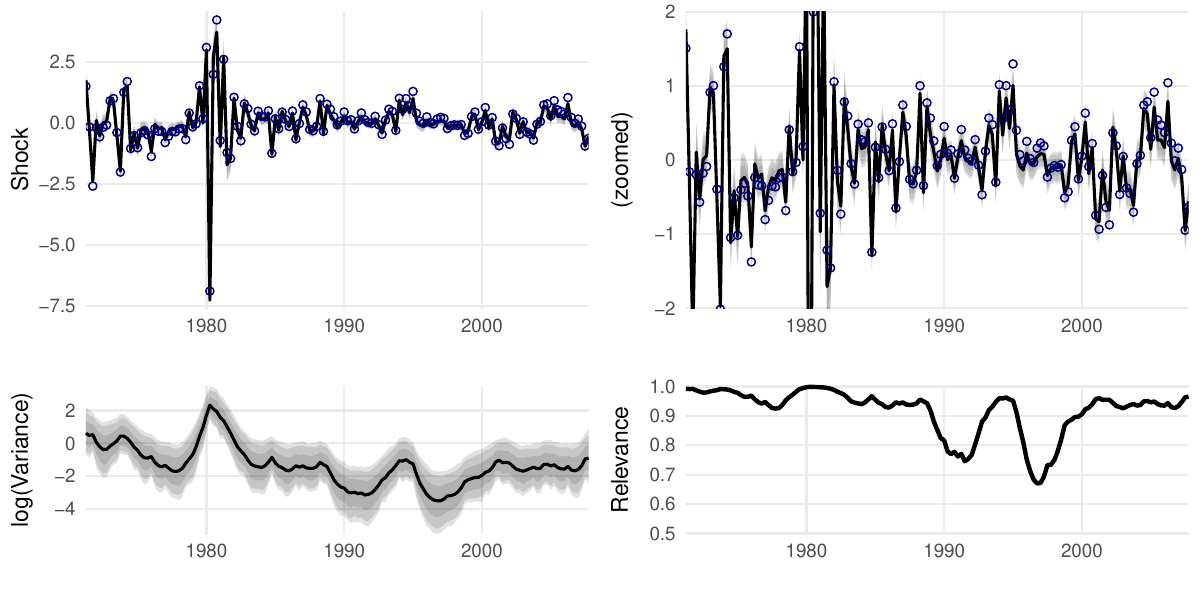}
    \caption{Estimated monetary policy shocks, their time-varying log-variance, and resulting relevance statistic. The gray shaded area marks the 68/90/95 percent credible set alongside the median (solid black line) and observed instrument (blue circles).}
    \label{fig:irf_mp_shocks}
\end{figure}

\begin{figure}[!t]
    \includegraphics[width=\textwidth]{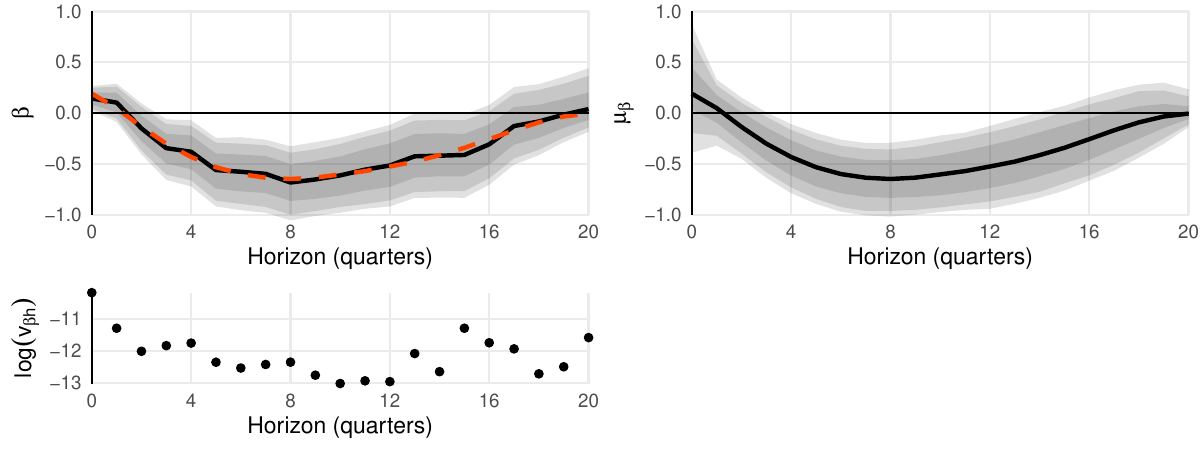}
    \caption{Estimated dynamic effects of a monetary policy shock on real GDP for SU-LP estimated with a GP prior and NG shrinkage. The gray shaded area marks the 68/90/95 percent credible set alongside the median (solid black line). The dashed red line is the posterior median of the GP prior.}
    \label{fig:irf_mp_main}
\end{figure}

\begin{figure}[t!]
    \begin{subfigure}[t]{0.24\textwidth}
    \caption{SU-LP}
    \includegraphics[width=\textwidth]{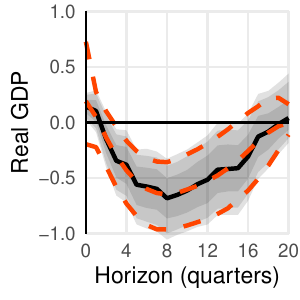}
    \end{subfigure}
    \begin{subfigure}[t]{0.24\textwidth}
    \caption{SU-LP (\textit{flat})}
    \includegraphics[width=\textwidth]{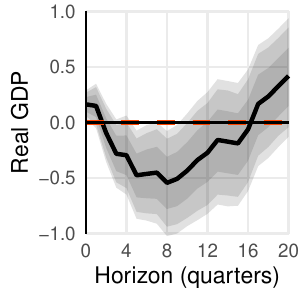}
    \end{subfigure}
    \begin{subfigure}[t]{0.24\textwidth}
    \caption{LP (\textit{default})}
    \includegraphics[width=\textwidth]{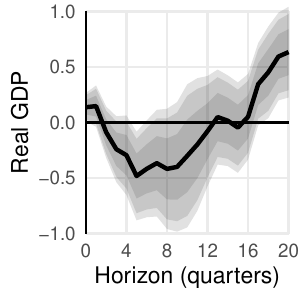}
    \end{subfigure}
    \begin{subfigure}[t]{0.24\textwidth}
    \caption{LP (\textit{smooth})}
    \includegraphics[width=\textwidth]{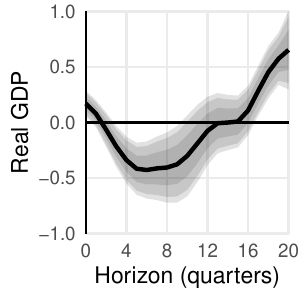}
    \end{subfigure}
    \caption{Comparison of estimated dynamic effects across different LP implementations. The gray shaded area marks the 68/90/95 percent credible set alongside the median (solid black line) for Bayesian implementations (and analogous confidence intervals and mean estimates for classical implementations). The dashed red lines are the posterior median of the GP prior alongside a 90 percent credible set.}
    \label{fig:irf_mp_comparison}
\end{figure}

The estimated shocks are shown in Fig. \ref{fig:irf_mp_shocks}, and the corresponding IRFs are shown in Figs. \ref{fig:irf_mp_main} (main specification) and \ref{fig:irf_mp_comparison} (comparison of alternative approaches). First, turning to the estimated shocks in Figure \ref{fig:irf_mp_shocks}, we plot the estimated shocks, associated posterior intervals, and the original instruments (depicted by dots) in the top two panels. Not surprisingly, the shock and instrument show clear heteroskedasticity, in particular around 1980 and the Volcker disinflation. This can also clearly be seen from our estimate of the volatility of the monetary shock in the lower left panel.  

The right upper panel zooms in on the estimated shock to make it clear that our estimates do not just simply replicate the instrument, but there is a meaningful measurement error. Finally, given the time-varying standard deviation of our monetary shock, we can assess how the relevance of the instrument (as defined above) changes over time. We find that the relevance is generally high, but falls in the 1990s. Nonetheless, we find that instrument relevance is not an issue in this application. Furthermore, note that a common complaint about the \cite{romer2004new} shocks is that they are autocorrelated. Our approach controls for this aspect by including lags of all relevant variables in the measurement equation for the instrument.

Next, we turn to impulse responses. Figure \ref{fig:irf_mp_main} plots the estimated impulse response in the upper left panel along with the GP posterior median in red dashes. The figure shows that output declines with a lag to a contractionary monetary shock, with a (negative) peak effect after around 8 quarters.  

We see that for this application, there is no meaningful difference between our estimated posterior of the impulse response and the GP (i.e., smoothed) counterpart, at least as far as the posterior median goes. The upper right panel then plots the GP posterior with error bands. Here we see that there are some minor differences at long horizons for the posterior bands, but nothing that changes the economic interpretation. The similarity is not surprising given our estimates of the variance of the difference between the posterior GP and the actual impulse response, which is plotted on a log scale in the lower panel. This result is application-specific, and for other data sets the estimated impulse response could be substantially less smooth than the GP posterior, which is smooth by construction.

Finally, we turn to a comparison of our approach with other alternatives. In particular, we compare the same set of models as in the previous section. Figure \ref{fig:irf_mp_comparison} plots our benchmark impulse response (leftmost panel), the LP from a version with a flat prior on $\bm \mu$, standard OLS estimation with HAC standard errors, and the \cite{barnichon2019impulse} LP estimator based on cross-validation. 

The patterns we find are very similar to those obtained using the simulated data. Our benchmark approach yields an impulse response that returns to zero after $20$ quarters, whereas the other alternatives show a positive response of GDP after $20$ quarters (zero is, however, contained in the $90$ percent credible set for the flat prior after $20$ quarters, whereas the last two specifications imply a significantly positive response after $20$ quarters). 

The transitory nature of the responses under the SU-LP model is mostly driven by the specification of the kernel that implies more shrinkage towards zero for higher-order responses in conjunction with the GL prior, effectively getting rid of the counter-intuitive result that output increases after around $16$ quarters. Notice, however, that this does not generally come at the expense of introducing strong biases. If the data suggests non-zero responses for longer horizons, our shrinkage prior would attribute little weight to the information arising from the GP piece. This interpretation depends partly on the long-horizon shrinkage implied by the benchmark kernel; the figure should therefore be read as combining information in the data with an explicit prior preference for transitory responses. Applications in which permanent effects are plausible would require a different prior mean, a different kernel, or a separate empirical design; we do not study that case here. 

\FloatBarrier

\subsection{Monetary and fiscal policy in the US}
Next, we estimate the responses to US monetary policy and government (defense) spending shocks in a joint framework, with the same setup of target (real GDP) and control variables as above. We again use the Romer and Romer monetary shocks and rely on a variant of the shocks of \citet{fisher2010using} to capture fiscal shocks (i.e., we have two shocks of interest, each with a single associated instrument). Another instrument for fiscal shocks that we use is developed by \citet{ben2017chronicle}. Our sample again ranges from 1970Q1 to 2007Q4. 

\begin{figure}[!ht]
    \begin{subfigure}[t]{\textwidth}
    \caption{\citet{fisher2010using} shocks}
    \includegraphics[width=\linewidth]{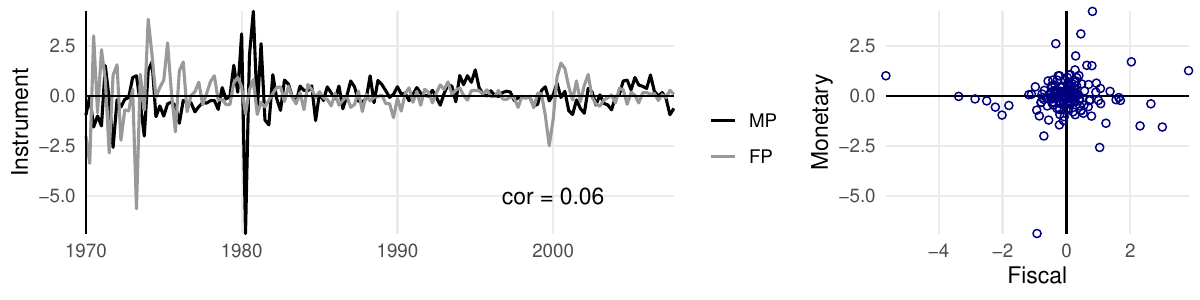}
    \end{subfigure}
    \begin{subfigure}[t]{\textwidth}
    \caption{\citet{ben2017chronicle} shocks}
    \includegraphics[width=\linewidth]{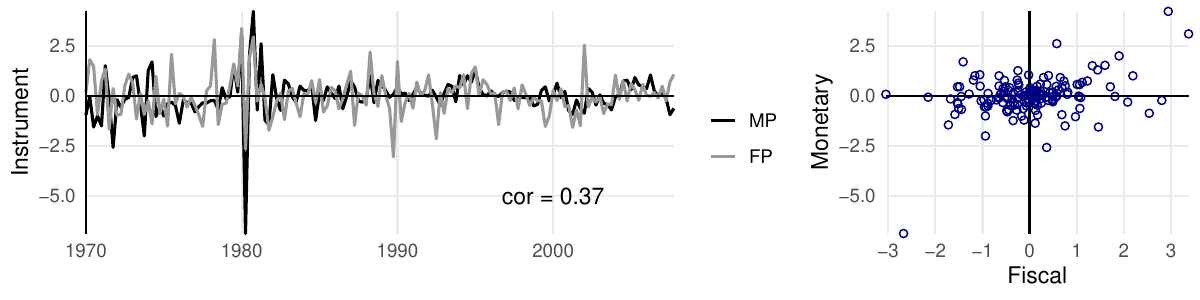}
    \end{subfigure}
    \caption{Time series and scatter plots of the standardized instruments, monetary policy (MP) and two variants for fiscal policy (FP); ``cor'' is the correlation coefficient.}
    \label{fig:data_jointinstruments}
\end{figure}

We use two fiscal shock instruments to highlight the effects of different modeling choices when it comes to taking into account the correlation between instruments, as the \citet{fisher2010using} instrument is barely correlated with our monetary shock instrument (a correlation of $0.06$), while \citet{ben2017chronicle} has a moderate unconditional correlation ($0.37$), as depicted in Figure \ref{fig:data_jointinstruments}, which shows both the time series and scatter plots of these various instruments and their correlation. 

To be clear, we do not mean this as a criticism of either instrument. A meaningful correlation can, for example, imply that there is (unexpected) coordination between monetary and fiscal policy that researchers should take into account --- this would be our scenario where there is a model-implied common component.

\begin{figure}[!ht]
    \begin{subfigure}[t]{0.49\textwidth}
    \caption{\citet{fisher2010using} shocks}
    \includegraphics[width=\textwidth]{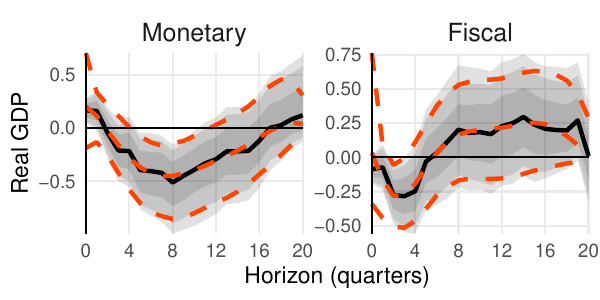}
    \end{subfigure}
    \begin{subfigure}[t]{0.49\textwidth}
    \caption{\citet{ben2017chronicle} shocks}
    \includegraphics[width=\textwidth]{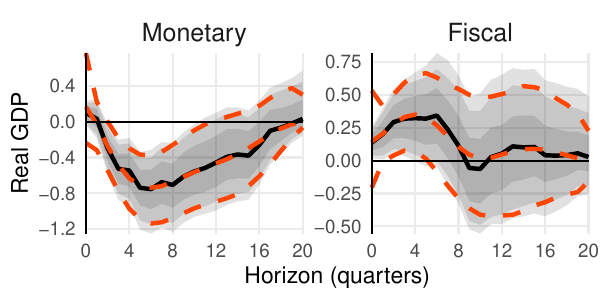}
    \end{subfigure}
    \caption{Comparison of estimated dynamic effects when assuming correlated measurement errors of monetary and fiscal shock instruments. The gray shaded area marks the 68/90/95 percent credible set alongside the median (solid black line). The dashed red lines are the posterior median of the GP alongside a 90 percent credible set.}
    \label{fig:irf_mpfp_main}
\end{figure}

We assume that both instruments measure the true shocks with error, and infer the latter as heteroskedastic latent states. We first account for the possibility of correlated measurement errors, not allowing for a model-implied common component across fiscal and monetary policy. The results are shown for the GP-NG version and a flat prior in Figure \ref{fig:irf_mpfp_main}. The impulse response to a monetary policy shock is largely unchanged across the two specifications, whereas the two fiscal instruments identify different impulse responses to a fiscal shock. \cite{ben2017chronicle} leads to a short-term increase in GDP up to 8 quarters, while the response to a \cite{fisher2010using} fiscal shock only leads to a substantial increase in GDP after 8 quarters (albeit with wider error bands).

How important is it to jointly model impulse responses in LPs and allow for correlation between the instruments to understand the effects of monetary shocks? To analyze this question, we now jointly model the responses to fiscal and monetary shocks. Throughout we use two specifications, keeping the monetary instrument the same, but using either fiscal instrument described above. 

\begin{figure}[t]
    \includegraphics[width=1\textwidth]{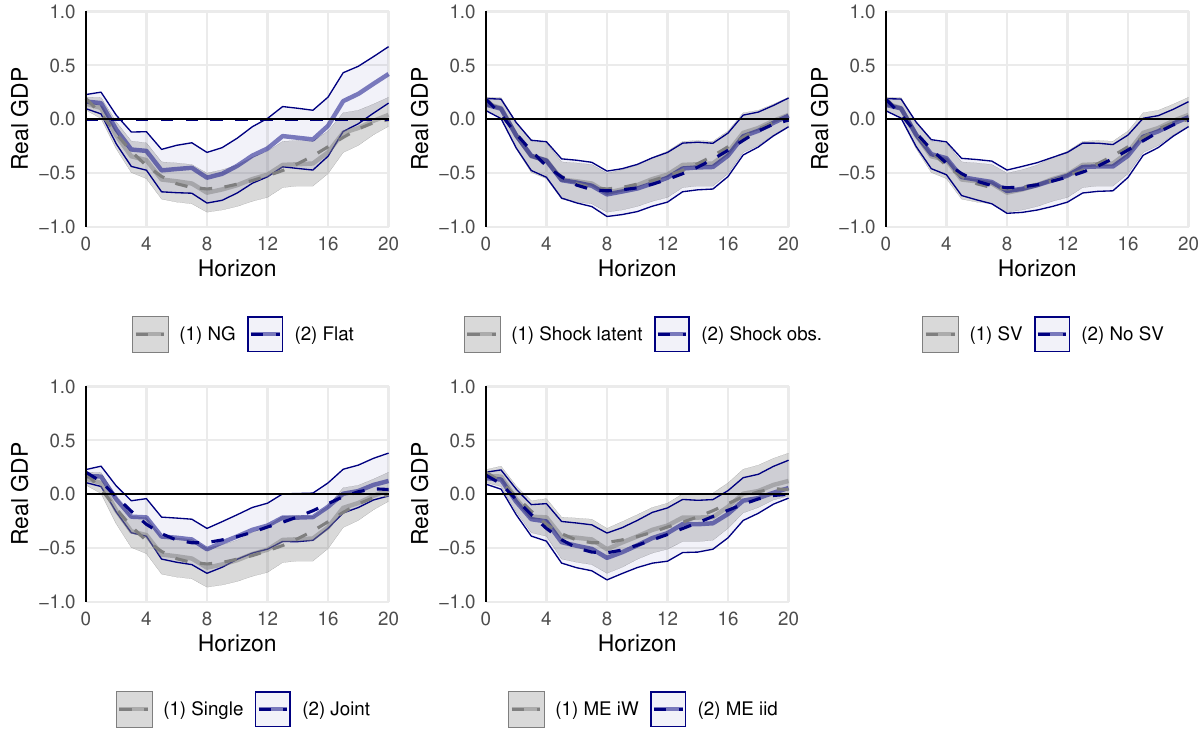}
    \caption{Comparing the response of real GDP to the monetary shock under different model specifications. Solid lines are the posterior median alongside the 68\% credible set, and the dashed lines are the smooth GP estimate. The top row shows the single-shock baseline specification (estimated with GP and NG prior, shocks with iid measurement error, and stochastic volatility) in gray. The blue variants by panel vary the indicated complication holding everything else fixed. The lower panels show a comparison to the joint-shock (including the fiscal shock of \citealp{fisher2010using}) case (left panel), and correlated versus independent measurement errors of the instruments (right panel). }
    \label{fig:irf_mp_comp_main}
\end{figure}

Figure \ref{fig:irf_mp_comp_main} plots the response to a monetary shock for various specifications. The upper row shows impulse responses when we only include the monetary instrument as in the previous section, whereas the lower row includes both the \cite{romer2004new} monetary shock instrument and the \cite{fisher2010using} fiscal instrument. The gray IRF is our benchmark from the previous section (except for the second panel on the bottom row, where we compare two cases with two instruments). 

It turns out that many specification choices in the single-instrument case only have very minor effects, with the exception of using a flat prior (top row, first panel). As compared to the informative NG prior, we find that medium-term responses (between 6 and 16 quarters) are stronger whereas shrinkage kicks in afterward, leading to longer-run responses that are increasingly centered on zero.

Treating the shock as observed versus latent or explicitly modeling stochastic volatility of the shock is inconsequential in this application. Introducing the fiscal shock has an effect (bottom row) even though the correlation between the \cite{romer2004new} instrument and the \cite{fisher2010using} instrument is low. Note that here we model the correlation between instruments as coming from the measurement error, a defensible assumption given the low correlation between the \cite{romer2004new} instrument and the \cite{fisher2010using} instrument. 

\begin{figure}
    \begin{subfigure}[t]{0.49\textwidth}
    \caption{\citet{fisher2010using} shocks}
    \includegraphics[width=0.49\textwidth]{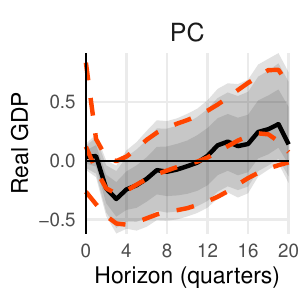}
    \includegraphics[width=0.49\textwidth]{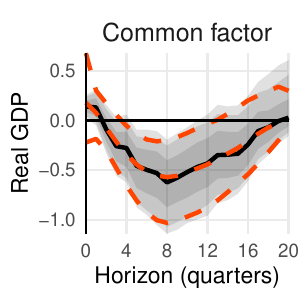}
    \end{subfigure}
    \begin{subfigure}[t]{0.49\textwidth}
    \caption{\citet{ben2017chronicle} shocks}
    \includegraphics[width=0.49\textwidth]{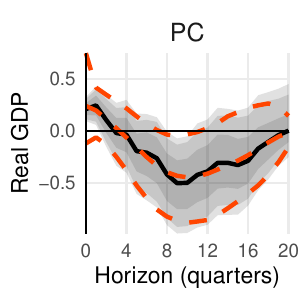}
    \includegraphics[width=0.49\textwidth]{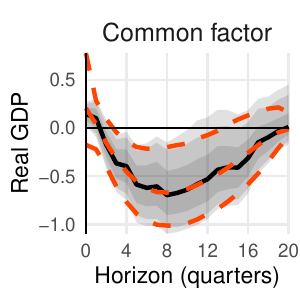}
    \end{subfigure}
    \caption{Estimating responses to the joint component of the monetary policy and fiscal shocks. ``PC'' extracts the first principal component from the instruments and includes it as a single instrument; ``Common factor'' estimates the joint component in our unified framework.}\label{fig:irf_jointshock}
\end{figure}

Next, we instead allow for a model-implied common component that influences both unexpected movements in monetary and fiscal policy, which opens the door for us to also use the \cite{ben2017chronicle} instrument. Figure \ref{fig:irf_jointshock} shows the estimated responses to the common component for our two fiscal instruments (keeping the monetary instrument the same across the two specifications). For both sets of instruments, we show results where we either ex-ante estimate the common component via principal components (the plots labeled ``PC'') or estimate it jointly with all other parameters (labeled ``Common factor''). Our interpretation is that these responses to the common component resemble the responses to monetary policy shocks we have shown above, with the slight exception of the PC case with the \cite{fisher2010using} instrument, where the qualitative pattern is similar, but the response is smaller in magnitude and the timing is somewhat different. 

We interpret this as suggestive, within this specification, that part of the effects that we usually attribute to monetary policy shocks might be attributable to a common fiscal and monetary component. A common component may also capture shared policy movements around recessions or other omitted macroeconomic conditions, so we interpret these estimates descriptively. The PC response and the common factor response are very similar for the \cite{ben2017chronicle} instrument, which is probably not surprising given the strong correlation between that instrument and the \cite{romer2004new} instrument. Our takeaway is that researchers should use the common component specification if there is a substantial correlation between instruments, but otherwise capturing a small correlation of instruments via correlation in measurement errors should be the benchmark choice.

\section{Conclusions}\label{sec:conclusions}
We have presented a general framework for inference in LPs that exploits advantages of the Bayesian toolkit. Our framework allows for multiple mismeasured shocks, can estimate a common shock that partially determines commonly used instruments in macroeconomics, allows for flexible, yet parsimonious regularization, requires minimal tuning of prior hyperparameters, models cross-horizon covariance in a stacked LP system, and automatically takes into account missing data. Relative to existing work, our approach delivers superior coverage probabilities in realistic environments. Finally, estimation is fast, removing one of the commonly cited advantages of standard frequentist inference in LPs.

{\setstretch{1.15}\putbib}
\end{bibunit}\clearpage

\begin{appendices}

\begin{center}
{\sffamily\Large\textbf{Online Appendix\\\huge\titletext}}
\end{center}

\setcounter{page}{1}
\setcounter{section}{0}
\setcounter{equation}{0}
\setcounter{footnote}{0}

\renewcommand\thesection{\Alph{section}}
\renewcommand\theequation{\Alph{section}.\arabic{equation}}
\renewcommand\thefigure{\Alph{section}.\arabic{figure}}
\renewcommand\theequation{\Alph{section}.\arabic{equation}}
\begin{bibunit}

\section{Technical Appendix}\label{app:technical}
\subsection{Posterior distributions and sampling algorithm}\label{sec:algorithm}
Let $\tilde{\bm{y}} = \bm{y} - \bm{Z}\bm{\gamma}$ be a vector of residuals. The posterior distribution of the vector of impulse responses is given by:
\begin{align}
    \bm{\beta}|\bm{y},\bm{\gamma},\bm{\Sigma}_u &\sim \mathcal{N}\left(\overline{\bm{\mu}}_\beta,\overline{\bm{V}}_\beta\right),\label{eq:irf_post}\\
    \overline{\bm{V}}_\beta &= \left(\bm{X}'\bm{\Sigma}^{-1} \bm{X} + \bm{V}_{\beta}^{-1}\right)^{-1},\nonumber\\
    \overline{\bm{\mu}}_\beta &= \overline{\bm{V}}_\beta \left(\bm{X}'\bm{\Sigma}^{-1}\tilde{\bm{y}} + \bm{V}_{\beta}^{-1}\bm{\mu}_\beta\right).\nonumber
\end{align}
We may rewrite the posterior expectation using the Woodbury matrix identity as:
\begin{align*}
    \mathbb{E}(\bm{\beta}|\bm{y},\bm{\gamma},\bm{\Sigma}_u) &= (\bm{X}'\bm{\Sigma}^{-1} \bm{X} + \bm{V}_{\beta}^{-1})^{-1}\bm{V}_{\beta}^{-1}\bm{\mu}_\beta + \left(\bm{I}_H - (\bm{X}'\bm{\Sigma}^{-1} \bm{X} + \bm{V}_{\beta}^{-1})^{-1}\bm{V}_{\beta}^{-1}\right)\hat{\bm{\beta}},\\
    &= \bm{W} \bm{\mu}_\beta + (\bm{I}_H - \bm{W})\hat{\bm{\beta}},
\end{align*}
where $\bm{W} = (\bm{X}'\bm{\Sigma}^{-1} \bm{X} + \bm{V}_{\beta}^{-1})^{-1}\bm{V}_{\beta}^{-1}$ are weights determined by the relative informativeness of the data and the prior, and $\hat{\bm{\beta}} = (\bm{X}'\bm{X})^{-1}\bm{X}'\tilde{\bm{y}}$ takes the form of the least squares estimator with the controls partialed out. That is, the posterior expectation is a convex combination of prior moments and data information.

For sampling the high-dimensional vector associated with the controls, it is convenient to rewrite the likelihood given in Equation (\ref{eq:SULP-lik}). Define $\bm{\mathrm{X}}$ of size $T \times n_x$, $\bm{\mathrm{Y}}$ is $T\times H$, and $\bm{\mathrm{Z}}$ is $T\times k$, with $t$th rows $\bm{x}_t'$, $\bm{y}_t'$ and $\bm{z}_t'$, respectively, then we may write:
\begin{equation*}
    \bm{\mathrm{Y}} = \bm{\mathrm{X}}\bm{B} + \bm{\mathrm{Z}}\bm{\Gamma} + \bm{U}, \quad \text{vec}(\bm{U})\sim\mathcal{N}(\bm{\Sigma}_u\otimes\bm{I}_T),
\end{equation*}
where $\bm{\gamma} = \text{vec}(\bm{\Gamma})$ where $\bm{\Gamma}$ is $k\times H$. The posterior distribution is:
\begin{align}
    \bm{\gamma}|\bm{Y},\bm{B},\bm{\Sigma}_u &\sim \mathcal{N}\left(\text{vec}\left(\overline{\bm{M}}_\gamma\right),\bm{\Sigma}_u \otimes \overline{\bm{V}}_\gamma\right),\label{eq:controls_post}\\
    \overline{\bm{V}}_\gamma &= (\bm{\mathrm{Z}}'\bm{\mathrm{Z}} + \bm{V}_\gamma^{-1})^{-1},\nonumber\\
    \overline{\bm{M}}_\gamma &= \overline{\bm{V}}_\gamma(\bm{\mathrm{Z}}'(\bm{\mathrm{Y}} - \bm{\mathrm{X}}\bm{B}) + \bm{V}_\gamma^{-1}\bm{M}_\gamma),\nonumber
\end{align}
which is a $kH$-dimensional Gaussian distribution. The Kronecker structure in the posterior makes this amenable to a fast sampling algorithm that avoids computing the Cholesky factor of the $kH\times kH$-sized posterior covariance matrix \citep[see, e.g.,][for a recent discussion in the VAR context]{chan2020large}.

The posterior of the covariance matrix is inverse Wishart distributed:
\begin{equation}
    \bm{\Sigma}_u|\bm{y},\bm{\beta},\bm{\gamma} \sim \mathcal{W}^{-1}\left(s_0 + T, \bm{S}_0 + \bm{U}'\bm{U}\right).\label{eq:Sigu_post}
\end{equation}

In case we assume that our instruments are measured with errors, one may draw the shocks $x_t$ from their joint Gaussian distribution conditional on all other parameters of the model. The variance of the measurement errors can then either be sampled from their well-known inverse Gamma $\sigma_{\nu}^2|\bullet\sim\mathcal{G}^{-1}(a_{\sigma\nu} + T/2, b_{\sigma\nu} + \sum_{t=1}^T \nu_t^2 / 2)$, in the case of a single instrument; or inverse Wishart distribution $\bm{\Sigma}_\nu|\bullet\sim\mathcal{W}^{-1}(s_{0\nu} + T, \bm{S}_{0\nu} + \sum_{t=1}^T \bm{\nu}_t'\bm{\nu}_t)$, when there are multiple instruments.

For updating the GP, under the prior given by Equation (\ref{eq:prior_GP}), the posterior distribution is:
\begin{equation}
    \bm{\mu}_{\beta} | \bm{\beta}, \bm{K}_\beta, \bm{V}_\beta \sim \mathcal{N}(\overline{\bm{\mu}}_{\beta}, \overline{\bm{K}}_\beta),\label{eq:post_GP}
\end{equation}
with mean $\overline{\bm{\mu}}_{\beta} = \bm{K}_\beta (\bm{K}_\beta + \bm{V_\beta})^{-1} \bm{\beta}$ and covariance matrix $\overline{\bm{K}}_\beta = \bm{K}_\beta - \bm{K}_\beta(\bm{K}_\beta + \bm{V}_\beta)^{-1}\bm{K}_\beta$.\footnote{In case we assumed a non-zero mean $\underline{\bm{\beta}} \neq \bm{0}$ in Equation (\ref{eq:prior_GP}), the posterior covariance in Equation (\ref{eq:post_GP}) would remain the same, but the mean then is given by $\overline{\bm{\mu}}_{\beta} = \underline{\bm{\beta}} + \bm{K}_\beta (\bm{K}_\beta + \bm{V_\beta})^{-1} (\bm{\beta} - \underline{\bm{\beta}})$.} 

To estimate the tuning parameters associated with the kernel of the GP, we note that the prior on $\bm{\beta}$ implicitly conditions on the hyperparameters $\varsigma,\xi$ because they determine $\bm{K}_\beta$. We now make this conditioning explicit, and state the conditional prior $p(\bm{\beta} | \bm{K}_\beta, \bm{V}_\beta)$ instead more concisely as $p(\bm{\beta} | \varsigma, \xi, \bullet)$, because it serves as a likelihood at the top hierarchy. The posterior distributions of interest are:
\begin{equation*}
    p(\varsigma | \bm{\beta}, \bullet) \propto p(\bm{\beta} | \varsigma, \bullet) p(\varsigma), \qquad p(\xi | \bm{\beta}, \bullet) \propto p(\bm{\beta} | \xi, \bullet) p(\xi),
\end{equation*}
neither of them is of a well-known form. We use random walk Metropolis-Hastings (RW-MH) steps to sample them. Let $\theta \in \{\varsigma,\xi\}$, and $\theta^{(c)}$ denotes the current ($c$) draw of a parameter and $\theta^{(\ast)}$ is a candidate draw; $q(\theta^{(\ast)}|\theta^{(c)})$ is a transition (proposal) density that states how one obtains a proposal conditional on the current state. The acceptance probability for the candidate draw is given by:
\begin{equation}
    \min\left(\frac{p(\bm{\beta} | \theta^{(\ast)}, \bullet) p(\theta^{(\ast)})}{p(\bm{\beta} | \theta^{(c)}, \bullet) p(\theta^{(c)})} \frac{q(\theta^{(c)}|\theta^{(\ast)})}{q(\theta^{(\ast)}|\theta^{(c)})} ,1\right).\label{eq:post_GPhyperMH}
\end{equation}

Note that the prior implied by Equations (\ref{eq:prior_Vbeta}) and (\ref{eq:prior_NG}) can be stated as $\beta_h | \lambda_h^2 \sim \mathcal{N}(\mu_{\beta h}, \lambda_h^2)$ with conditional distribution $\lambda_h^2|\tilde{\tau}^2 \sim \mathcal{G}(\vartheta_\lambda, \vartheta_\lambda \tilde{\tau}^2 / 2)$. This allows for deriving the conditional posteriors of these parameters, which are given by:
\begin{align}
    \lambda_h | \bullet &\sim \mathcal{GIG}\left(\vartheta_\lambda - 1/2, (\beta_h - \mu_{\beta h})^2, \vartheta_\lambda \tilde{\tau}^2\right),\label{eq:post_NG}\\
    \tilde{\tau}^2 | \bullet &\sim \mathcal{G}\left(a_\tau + \vartheta_\lambda H, b_\tau + \vartheta_\lambda / 2 \sum_{h=0}^{\tilde{H}}\lambda_{h}^2\right).\nonumber
\end{align}

\subsection{Vector autoregressive moving average (VARMA) DGPs}\label{subsec:varmadgp}
\subsubsection{Calibration of reduced form VAR parameters}
The variables included in the VAR($5$) that we use for calibration of the DGP paramaeters are real GDP (\texttt{GDPC1}), personal consumption expenditures (\texttt{PCECC96}), gross private domestic investment (\texttt{GPDIC1}), GDP deflator (\texttt{GDPCTPI}), hours worked for all workers (\texttt{HOABS}), all in annualized log-differences (FRED codes in parentheses); federal funds rate (\texttt{FEDFUNDS}) and Moody's seasoned BAA corporate bond yield (\texttt{BBA}). The parameters are estimated using standardized data for the sampling period from 1960Q1 to 2019Q4. 

For estimation, we rely on a conjugate Minnesota prior with hierarchically estimated hyperparameters, as in \citet{giannone2015prior}, and use the posterior median estimate for simulating our DGPs. 

\subsubsection{Companion form and impulse response function} 
The companion form of the VARMA processes we consider as DGPs is given by:
\begin{equation*}
    \bm{W}_t = \bm{F} \bm{W}_{t-1} + \bm{M}\bm{H}\bm{\epsilon}_t + \alpha T^{-\pi}\sum_{j = 1}^{\infty}\bm{M}\bm{A}_j\bm{H}\bm{\epsilon}_{t-j},
\end{equation*}
where $\bm{W}_t = (\bm{w}_t',\hdots,\bm{w}_{t-P+1}')'$, $\bm{F} = [\bm{\Phi}_1, \hdots, \bm{\Phi}_P; \bm{I}_{n(P-1)}, \bm{0}_{n(P-1)\times n}]$ is $nP \times nP$, $\bm{M} = (\bm{I}_n,\bm{0}_{n\times n(P-1)})'$. The importance of the MA part (and thus the degree of misspecification when using VARs for estimation) is a function of the sample size, governed by $\alpha T^{-\pi}$, see \citet{schorfheide2005var} for additional discussions. One may obtain $\bm{w}_{t+h|t-1} = \bm{M}'\bm{W}_{t+h}$ for $h = 1,2,\hdots,$ recursively:
\begin{equation*}
    \bm{M}'\bm{F}^{h+1}\bm{W}_{t-1} + \sum_{l=0}^h \bm{M}'\bm{F}^{l}\bm{M}\bm{H}\bm{\epsilon}_{t+h-l} + \alpha T^{-\pi} \sum_{l=0}^h\sum_{j=1}^{\infty} \bm{M}'\bm{F}^{l}\bm{M}\bm{A}_j\bm{H}\bm{\epsilon}_{t+h-l-j}.
\end{equation*}
The shock impact at $h = 0$ is given by $\bm{H}$, while the dynamic response functions for horizons $h = 1, 2, \hdots$ are given by:
\begin{equation*}
    \frac{\partial \bm{w}_{t+h}}{\partial \bm{\varepsilon}_t} = \bm{M}'\bm{F}^h\bm{M}\bm{H} + \alpha T^{-\pi} \sum_{k=1}^{h} \bm{M}'\bm{F}^{h-k}\bm{M}\bm{A}_k\bm{H},
\end{equation*}
which is obtained by noticing that only the time $t$ structural shocks are of interest. This significantly simplifies the infinite double sum. The $(i,j)$th element of this matrix is the IRF of the $i$th variable to the $j$th structural shock at horizon $h$, i.e., $\partial w_{it+h} / \partial \varepsilon_{jt} = \beta_{\ast}^{(h)}$, and the true IRF is $\bm{\beta}_{\ast} = (\beta_{\ast}^{(0)}, \beta_{\ast}^{(1)},\hdots,\beta_{\ast}^{(\tilde{H})})'$.

\begin{figure}[ht]
    \begin{subfigure}[t]{0.49\textwidth}
    \caption{$\alpha = 0$}
    \includegraphics[width=\textwidth]{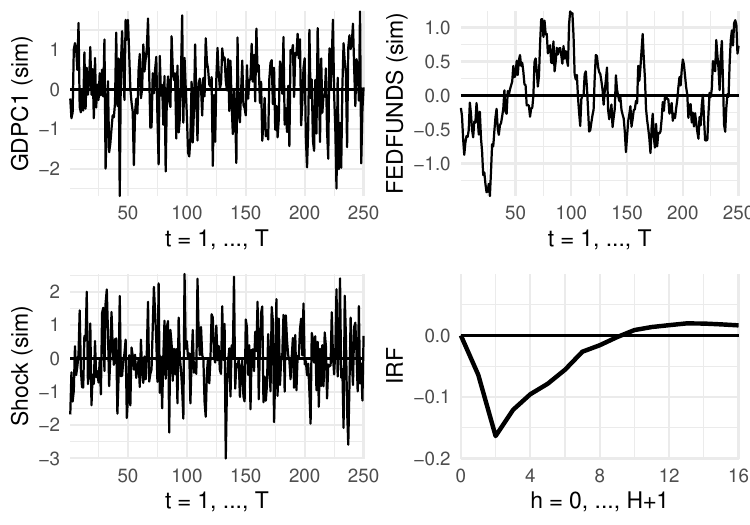}
    \end{subfigure}
    \begin{subfigure}[t]{0.49\textwidth}
    \caption{$\alpha = 2$}
    \includegraphics[width=\textwidth]{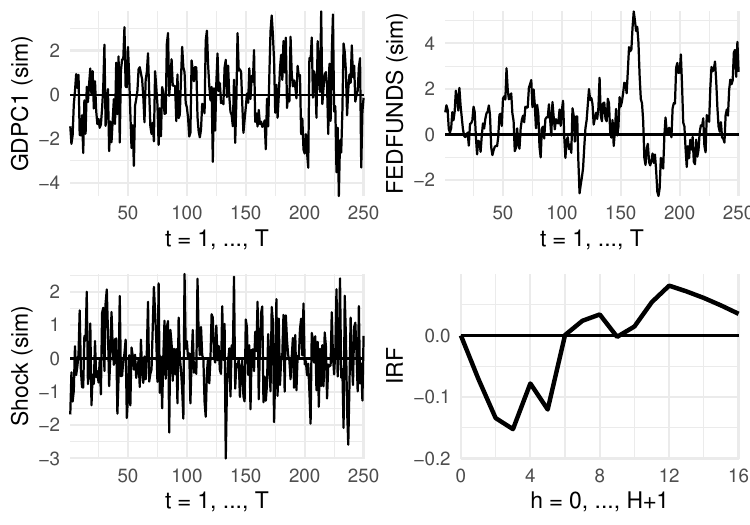}
    \end{subfigure}
    \caption{Realizations of selected variables and associated target impulse response function for across DGPs.}\label{fig:simdgpVARMA}
\end{figure}

Figure \ref{fig:simdgpVARMA} provides a realization of two DGPs for $T = 250$ and settings for $\alpha$, for the simulated target variable (\texttt{GDPC1}), policy instrument (\texttt{FEDFUNDS}) and the associated observed structural shock (\texttt{Shock}) alongside the true impulse response function (IRF).

\subsection{Forcing homogeneity of one-step ahead prediction errors}
Treating the one-step ahead prediction errors $\bm{e}$ as unobserved states that are linked to the $h$-specific innovations $\bm{u} = (\bm{u}_1',\hdots,\bm{u}_{T+H}')' = \text{vec}(\bm{U}')$ allows to write $\bm{u} = (\bm{I}_{T} \otimes \bm{Q}) \bm{\mathrm{S}}_e \bm{e}$. Here, $\bm{\mathrm{S}}_e$ is a $T\times (T+H)$ selection matrix of zeroes and ones that singles out the appropriate leads of $\bm{e}$. Rather than a full covariance matrix $\bm{\Sigma}_u$, this specification thus parameterizes its lower Cholesky factor $\bm{Q}$ instead.

A computationally efficient implementation can be achieved by assuming an approximate version of this measurement equation, which adds a small measurement error:
\begin{equation*}
    \bm{u} = (\bm{I}_{T} \otimes \bm{Q}) \bm{\mathrm{S}}_e \bm{e} + \bm{\varkappa}, \quad \bm{e} \sim \mathcal{N}(\bm{0}_{T+H},\sigma^2_e \bm{I}_{T+H}), \quad \bm{\varkappa} \sim \mathcal{N}(\bm{0}_{T+H},\mathfrak{o}^2\bm{I}_{T+H}),
\end{equation*}
where $\sigma^2_e$ is the variance of the (homogenized) one-step ahead prediction error, and $\mathfrak{o}^2 = 10^{-8}$ is set to a very small value. This representation can be used to sample $\bm{e}$ from its multivariate Gaussian posterior which takes a textbook form and is amenable to precision sampling. 

Given a draw of $\bm{e}$, one could update $\sigma^2_e$, and loop through horizons to update $\bm{Q}$ under conditionally conjugate priors, e.g., an inverse Gamma prior on $\sigma^2_e$ and independent Gaussian priors on the unrestricted elements of $\bm{Q}$, $q_{ij}$.

\clearpage
\section{Empirical Appendix}\label{app:empirical}
\subsection{Simulated data: DSGE-based DGP}
We simulate $T=250$ observations from the  \cite{smets2003estimated} dynamic stochastic general equilibrium (DSGE) model, and estimate the response of output to a monetary policy shock.  The structural parameters of the model are obtained along the lines of \cite{smets2003estimated} and implemented in the \texttt{R} package \texttt{gEcon} \citep[see][]{klima2015smets}.

\begin{figure}[ht]
    \begin{subfigure}[t]{0.24\textwidth}
    \caption{SU-LP}
    \includegraphics[width=\textwidth]{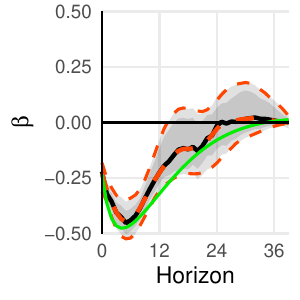}
    \end{subfigure}
    \begin{subfigure}[t]{0.24\textwidth}
    \caption{SU-LP (\textit{flat})}
    \includegraphics[width=\textwidth]{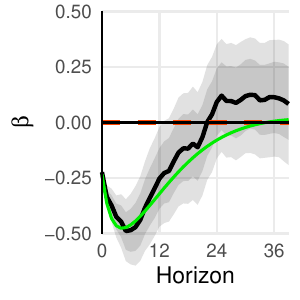}
    \end{subfigure}
    \begin{subfigure}[t]{0.24\textwidth}
    \caption{LP (\textit{default})}
    \includegraphics[width=\textwidth]{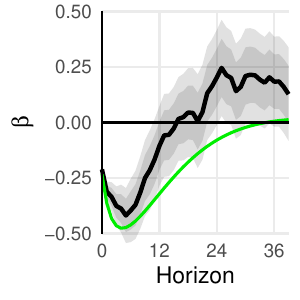}
    \end{subfigure}
    \begin{subfigure}[t]{0.24\textwidth}
    \caption{LP (\textit{smooth})}
    \includegraphics[width=\textwidth]{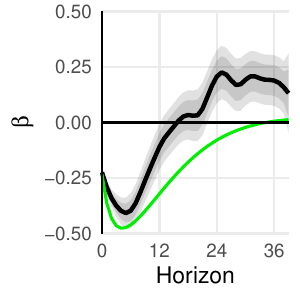}
    \end{subfigure}
    \caption{Comparison of estimated dynamic effects across different LP implementations. The gray shaded area marks the 68/90/95 percent credible set alongside the median (solid black line), and analogous confidence intervals and mean estimates for classical versions. The dashed red lines are the posterior median of the GP prior alongside a 90 percent credible set. The green line is the true impulse response function.}
    \label{fig:irf_sim_comparison}
\end{figure}

Figure \ref{fig:irf_sim_comparison} plots the output responses for our benchmark case with a GP prior, a flat prior in the impulse response coefficients, OLS with HAC standard errors, as well as the smoothed LPs of \cite{barnichon2019impulse}. For our approach, we also plot the posterior for the Gaussian process in red, which we can interpret as an estimate of a smoothed LP.  

All four methods considered produce declines in output in response to contractionary monetary policy shocks, with a peak effect after around six periods. What sets our benchmark approach apart from the others is that via the smoothing implied by the GP, we do fit the longer horizon response much better than the alternatives, which implies an erroneous rise in GDP after approximately two years. With flat priors, our approach is reasonably close to the OLS-based IRFs, but closer to the truth than the frequentist alternatives. Note that these two approaches differ, apart from the underlying statistical paradigm and the presence of priors, in their horizon-specific sample sizes.

\begin{figure}[t]
    \centering
    \includegraphics[width=0.4\textwidth]{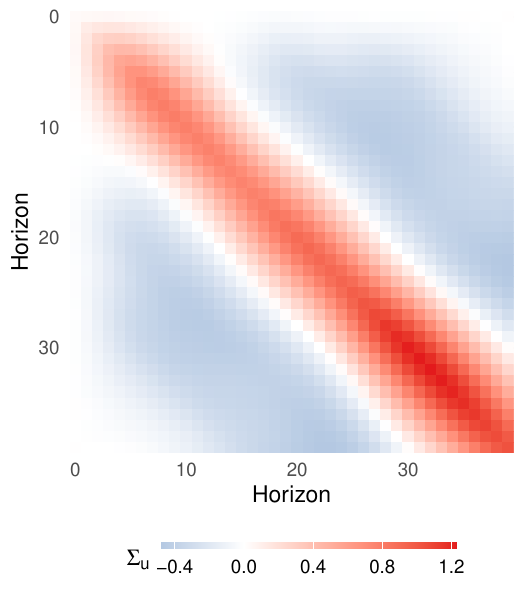}
    \includegraphics[width=0.4\textwidth]{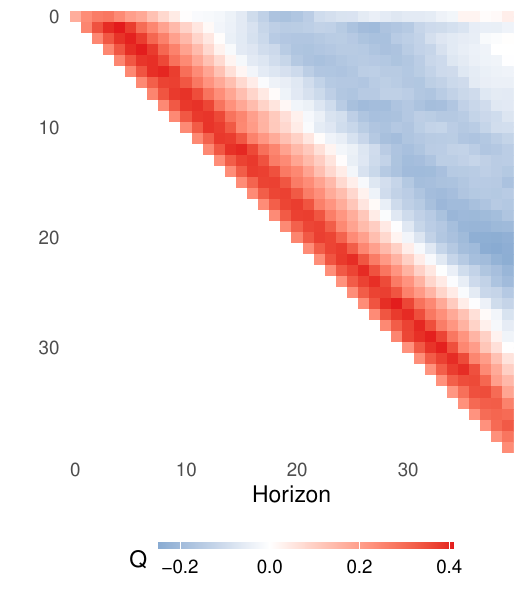}
    \caption{Posterior median estimate of the covariance structure $\bm{\Sigma}_u$ across horizons $h$ alongside implied MA($h$) parameters in $\bm{Q}$.}
    \label{fig:irf_sim_Hcov}
\end{figure}

Figure \ref{fig:irf_sim_Hcov} plots the estimated (median) posterior covariance matrix of the forecast errors along with its Cholesky factor. We show these results to highlight that indeed there seems to be a pattern in this matrix that is broadly consistent with the results from our earlier examples. More generally, our approach captures the comovement between forecast errors across horizons.

{\setstretch{1.2}\putbib}
\end{bibunit}

\end{appendices}

\end{document}